\documentclass[12pt,letterpaper]{article}
\usepackage[usenames, dvipsnames]{xcolor}
\usepackage{jcapmod}

\usepackage{verbatim}
\usepackage{mathrsfs}
\usepackage{cleveref}
\usepackage[shortlabels]{enumitem}

\usepackage{setspace,caption}

\usepackage{bbm}
\usepackage{graphicx}
\usepackage{bm}
\usepackage{multirow}

\usepackage{amsmath}
\usepackage{amssymb}
\usepackage{amsfonts}
\usepackage{mathtools}

\newcommand{\cH}{\mathcal{H}}

\newcommand{\cM}{\mathcal{M}}

\newcommand{\cO}{\mathcal{O}}
\newcommand{\cS}{\mathcal{S}}

\newcommand{\ud}{\mathrm{d}}
\newcommand{\ue}{\mathrm{e}}
\newcommand{\ui}{\mathrm{i}}

\newcommand{\R}{\mathbb{R}}
\newcommand{\Z}{\mathbb{Z}}
\newcommand{\C}{\mathbb{C}}

\newcommand{\ka}{\kappa}
\newcommand{\lam}{\lambda}
\newcommand{\Lam}{\Lambda}
\newcommand{\Ga}{\Gamma}
\newcommand{\ep}{\epsilon}
\newcommand{\ga}{\gamma}

\newcommand{\abs}[1]{\left\vert#1\right\rvert}
\newcommand{\nn}{\nonumber}

\newcommand{\Lstar}{L_{*}}
\newcommand{\lamstar}{\lam_{*}}
\newcommand{\rhostar}{\rho_{*}}
\newcommand{\Fock}{\cH_{\mathrm{Fock}}}
\newcommand{\BG}{\mathfrak{B}}
\newcommand{\eR}{\widetilde{R}}
\newcommand{\eD}{\widetilde{D}}
\newcommand{\QB}{Q_B}

\newcommand{\KBc}{\mathit{KBc}}

\newcommand{\bra}[1]{\langle #1|}
\newcommand{\ket}[1]{|#1\rangle}

\definecolor{cobalt}{RGB}{44, 98, 120}
\hypersetup{
  colorlinks,
  citecolor=Violet,
  linkcolor=cobalt,
  urlcolor=Blue}

\begin{document}

\begin{titlepage}
\setcounter{page}{1} \baselineskip=15.5pt \thispagestyle{empty}
\setcounter{tocdepth}{2}
\bigskip\

\vspace{1cm}
\begin{center}
{\large \bfseries Recursive-algebraic solution of the closed string tachyon vacuum equation}
\end{center}

\vspace{0.55cm}

\begin{center}
\scalebox{0.95}[0.95]{{\fontsize{14}{30}\selectfont Manki Kim$^{a,b}$\vspace{0.25cm}}}
\end{center}

\begin{center}
\vspace{0.15 cm}
{\fontsize{11}{30}
\textsl{$^{a}$Leinweber Institute for Theoretical Physics,
Stanford University, Stanford, CA 94305}\\[4pt]
\textsl{$^{b}$Physical Superintelligence PBC (current affiliation)}}\\
\vspace{0.25cm}
\vskip .5cm
\end{center}

\vspace{0.8cm}
\noindent
We develop a recursive algebraic framework for solving the closed string tachyon vacuum equation, derived from the hyperbolic recursion relations of F{\i}rat and Valdes-Meller. We restrict to the sector of zero-momentum Lorentz-scalar states. Lorentz symmetry ensures that this sector is closed under the equations of motion. In this sector, we introduce a seam-graded expansion and show that the equation is entirely algebraic at every order: the unknown at each grade enters only through point evaluations at the systolic length, so each grade reduces to a matrix inversion with no Fredholm equations. The expansion is formal; convergence in the multi-level system is the subject of ongoing work. This work was conducted with a publicly available version of Claude Code (Anthropic, Claude Opus 4.6). The complete research repository, including all computations, adversarial review logs, and the full human-AI collaboration history, is publicly available at \url{https://github.com/mk2427/csft-tachyon-vacuum}.

\vspace{3.1cm}

\noindent\today

\end{titlepage}
\tableofcontents\newpage

\section{Introduction}\label{sec:intro}

The fate of the closed string tachyon is one of the central open questions in string theory. Unlike the open string tachyon, whose condensation describes the decay of an unstable D-brane and is by now well understood ~\cite{Schnabl:2005gv,Okawa:2006vm,Erler:2009uj,Erler:2014eba,Kiermaier:2010cf,Murata:2011ex,FuchsKroyter:2006,FuchsKroyterPotting:2007,GaiottoRastelli:2005,MoellerTaylor:2004} (see~\cite{Erler:2019vhl} for a review), the closed string tachyon couples to the metric and dilaton: when it condenses, it is spacetime itself that is restructured. Building on conjectures of Sen, Yang and Zwiebach~\cite{Yang:2005rx} argued that the endpoint of tachyon condensation in the bosonic string is a ``nothing'' state with vanishing cosmological constant, in which the energy density of the perturbative vacuum is exactly canceled by the tachyon condensate (see also~\cite{Bergman:1999km,Erler:2022agw,Erler:2023emp}). An analytic construction of this vacuum would provide a nonperturbative definition of the closed string landscape and a concrete realization of background independence.

Recent years have seen significant progress in string field theory (for reviews, see~\cite{Sen:2024nfd,Erbin:2021smf,Maccaferri:2023vns}). Recent advances include a complete formulation of superstring field theories \cite{Sen:2015hia,Sen:2014dqa,Sen:2015hha,Sen:2015uaa,FarooghMoosavian:2019yke}, the D-instanton amplitudes \cite{Sen:2019qqg,Sen:2020cef,Sen:2021qdk,Sen:2021tpp,Sen:2021jbr,Eniceicu:2022xvk,Eniceicu:2022dru,Chakravarty:2022cgj,Alexandrov:2021shf,Alexandrov:2021dyl,Alexandrov:2022mmy,Agmon:2022vdj,Sen:2024zqr,Scheinpflug:2026vaw}, D0-brane recoil problem \cite{Sen:2025xaj}, study of string backgrounds without the exact CFT including Ramond-Ramond backgrounds \cite{Cho:2018nfn,Cho:2023mhw,Kim:2024dnw,Cho:2025coy,Frenkel:2025wko}, the boundary action in string field theory \cite{Stettinger:2024uus,Firat:2024kxq,Maccaferri:2025orz,Maccaferri:2025onc}, covariant phase space \cite{Cho:2023khj,Bernardes:2025zkj,Bernardes:2025zzu,Bernardes:2025uzg}, identification of closed string modes in open string field theory \cite{Choi:2026ljv}, and the introduction of hyperbolic string vertices~\cite{Costello:2019fuh,Cho:2019anu,Moosavian:2017qsp,Moosavian:2017sev,Firat:2021ukc,Firat:2023sld} which opened new avenues for constructing string vertices systematically.

Despite this progress, the closed string tachyon vacuum has resisted a deep understanding. The obstacle is structural: covariant closed string field theory is intrinsically nonpolynomial~\cite{SaadiZwiebach:1989,KugoSuehiro:1989}, with vertices $V_{g,n}$ at every order required by the BV master equation~\eqref{eqn:geometric BV}. There is no star product and no associative algebra, so the $\KBc$ subalgebra that solved the open string problem has no direct counterpart. Finding a closed-string analogue of the $\KBc$ algebra adapted to hyperbolic vertices is a central open problem: such an algebra could yield the grade-$0$ seed analytically and make the entire expansion algebraic end-to-end. Moreover, no consistent covariant cubic closed string field theory exists~\cite{Sonoda:1989wa}: the quartic and higher vertices are mandatory.


A breakthrough came with the work of F{\i}rat and Valdes-Meller~\cite{Firat:2024ajp}, who showed that the topological recursion of hyperbolic closed string field theory reduces the classical equation of motion to a single \emph{quadratic integral equation} (their eq.~(6.9)). The key insight is that in the hyperbolic formulation, the Mirzakhani recursion for Weil--Petersson volumes translates directly into a recursion for string vertices: all higher-order vertices $V_{0,n}$ are generated from the cubic vertex $V_{0,3}$ alone. The resulting equation involves only the cubic vertex and the twisted Mirzakhani kernels, and the string field $\Phi(L_1)$, encoding a self-consistency condition for the tachyon vacuum. All ingredients --- the hyperbolic three-vertex $V_{0,3},$ the twisted Mirzakhani kernels $\eR$ and $\eD,$ and the Fenchel--Nielsen $b$-ghost $\BG$ --- are known explicitly. As a proof of concept, F{\i}rat and Valdes-Meller solved the analogous equation in the stubbed bosonic quantum field theory and recovered the full nonperturbative solution exactly from cubic vertex data~\cite{Firat:2024ajp}.\footnote{See also \cite{Erler:2023emp} for an earlier work on the similar idea.}

In this paper, we restrict to the sector of zero-momentum Lorentz-scalar states --- the natural starting point and the setting of Yang and Zwiebach's analysis. Lorentz symmetry ensures that this sector is closed under the equations of motion: scalar states cannot source non-scalar fields through the cubic vertex. Whether the physical vacuum lies within this sector is an open question --- it may require non-scalar condensation or nontrivial spacetime profiles. In the general vertex, the string field enters with arbitrary spacetime derivatives, and the equation of motion is a nonlinear integro-differential equation in both the border length and spacetime. In the scalar sector, these derivative terms are absent, and the equation simplifies dramatically. We introduce a \emph{seam-graded algebra} on the space of string field configurations and show that the resulting grade-by-grade expansion of eq.~(6.9) is entirely algebraic at every order. A crucial feature is that the approach bypasses the explicit construction of the higher string vertices $V_{0,4},$ $V_{0,5},\ldots,$ which are notoriously difficult to compute~\cite{Belopolsky:1994sk,Moeller:2004yy}: the Mirzakhani recursion encodes all higher-vertex contributions implicitly through the twisted kernels $\eR$ and $\eD,$ so the only vertex data needed is the cubic $V_{0,3}.$ 

The seam-graded expansion decomposes the quadratic integral equation into a sequence of manageable pieces: grade~$0$ keeps only the cubic vertex, grade~$1$ adds the quartic vertex (a single integral), grade~$2$ adds the quintic (a double integral), and so on. At every grade $n\geq 1,$ the unknown $\Phi_n$ enters only at the threshold length $L,$ reducing what would naively be a Fredholm integral equation to a matrix equation. The full string field is reconstructed by summing the grade-by-grade corrections.

The paper is organized as follows. \S\ref{sec:review} provides a self-contained review of the F{\i}rat--Valdes-Meller formulation, covering the structure of the closed string field theory action, the hyperbolic three-string vertex and its explicit local coordinates, the Fenchel--Nielsen $b$-ghost, the topological recursion that builds all higher vertices from the cubic one, and the review of the quadratic integral equation that is the central object of this paper. \S\ref{sec:seam-algebra} introduces the seam grading, works out the grade-$0,$ $1,$ and $2$ equations in detail, proves that the algebraic structure persists at every grade above the seed, and extends the formalism to the full Fock space by deriving the multi-level linearized operator $M = 2JB$ (\S\ref{sec:M-equals-2J}). \S\ref{sec:numerical} presents the numerical results: grade-$0$ solutions at level truncations up to $(0{+}2{+}4),$ the eigenvalue spectrum of $M(\Lstar),$ the grade-$1$ source computation and its analytic continuation via three independent methods. \S\ref{sec:comparison-YZ} compares our results with the Yang--Zwiebach level-truncation analysis, explaining how the seam-graded expansion recovers and extends their findings. Finally, \S\ref{sec:discussion} discusses convergence, the role of special functions and Barnes integrals, resummation strategies, the quantum extension, and future directions.

\section{Review of the FVM formulation}\label{sec:review}

Closed string field theory has a long history, and its full construction is technically demanding. For the reader's convenience, we collect here a self-contained account of the elements that are directly relevant to this paper. We begin with the motivations and structure of the closed string field theory action (\S\ref{sec:review-csft}), then describe the hyperbolic three-string vertex (\S\ref{sec:review-vertex}), the Fenchel--Nielsen b-ghost insertion (\S\ref{sec:review-bghost}), the topological recursion that builds all higher vertices from the cubic one alone (\S\ref{sec:review-recursion}), and finally the quadratic integral equation that is the main object of study in this paper (\S\ref{sec:review-eq69}). Throughout, we follow the conventions of F{\i}rat and Valdes-Meller~\cite{Firat:2024ajp}; for general reviews of string field theory, see~\cite{Zwiebach:1992ie,Erler:2019vhl,deLacroix:2017lif,Sen:2024nfd,Erbin:2021smf,Maccaferri:2023vns}.

\subsection{Closed string field theory}\label{sec:review-csft}

String perturbation theory, as usually formulated, is an on-shell framework: scattering amplitudes are defined as integrals over the moduli space of Riemann surfaces, with physical vertex operators inserted at the punctures. This framework is powerful for computing S-matrix elements, but it has well-known limitations. Off-shell quantities are not directly accessible~\cite{Sen:2015hia}. Background independence is obscured, since the theory is formulated with a fixed reference CFT on the worldsheet. And questions about the nonperturbative structure of string theory, including vacuum transitions and tachyon condensation, cannot even be posed in the on-shell language.

Closed string field theory was constructed precisely to overcome these limitations~\cite{Zwiebach:1992ie}. Its dynamical variable is a string field $\Psi,$ living in the restricted state space $\widehat{\cH}\subset\Fock,$ defined by the conditions
\begin{equation}
    b_0^-\ket{\Psi} = 0\,,\qquad L_0^-\ket{\Psi} = 0\,,\label{eqn:string field constraints}
\end{equation}
where $b_0^{\pm} = b_0 \pm \bar{b}_0$ and $L_0^{\pm} = L_0 \pm \bar{L}_0.$ The first condition projects onto states annihilated by $b_0^-,$ and the second is the level-matching constraint, guaranteeing invariance under rigid rotation of the closed string. These are kinematic constraints that define the physical state space, not gauge conditions. The BPZ inner product on $\widehat{\cH}$ is
\begin{equation}
    \langle A,\, B\rangle = \bra{A}\,c_0^-\,\ket{B}\,,\label{eqn:BPZ inner product}
\end{equation}
where $\bra{A}$ denotes the BPZ conjugate of $\ket{A}$ (defined by the conformal map $z\mapsto 1/z$), and the insertion of $c_0^- = \tfrac{1}{2}(c_0-\bar{c}_0)$ is necessary to obtain a nondegenerate pairing on the $b_0^-=0$ subspace. The ghost number anomaly on the sphere demands total ghost number $6$; since $c_0^-$ carries ghost number $+1,$ the inner product is nonvanishing precisely when $\mathrm{gh}(A) + \mathrm{gh}(B) = 5.$

Zwiebach's construction~\cite{Zwiebach:1992ie} assembles the full quantum action from the geometry of moduli space. The key idea is that string interactions arise from the decomposition of the moduli space $\cM_{g,n}$ of genus-$g$ Riemann surfaces with $n$ punctures into vertex regions and Feynman regions, covering $\cM_{g,n}$ exactly once. Each vertex is specified by a choice of local coordinates around every puncture, which determines how the CFT correlator on the surface is evaluated. The resulting action takes the form
\begin{equation}
    S[\Psi] = \frac{1}{2}\bra{\Psi}c_0^-\QB\ket{\Psi} + \sum_{\substack{g,n \\ 2g-2+n>0}}\frac{\hbar^g\,\ka^{2g-2+n}}{n!}\langle V_{g,n}|\Psi\rangle^{\otimes n}\,,\label{eqn:CSFT action}
\end{equation}
where $\QB$ is the BRST operator, $\ka$ is the string coupling constant, and the elementary vertices $\langle V_{g,n}|$ are multilinear maps on the Fock space. The first term is the standard free kinetic term; the remaining terms form an infinite tower of interactions --- cubic, quartic, and so on through all genera and puncture numbers. This \emph{nonpolynomial} structure is a fundamental distinction from open string field theory, where the Witten action is cubic. There is no consistent covariant, level-matched, cubic closed string field theory~\cite{Sonoda:1989wa}, and Zwiebach's construction is the minimal one that achieves consistency.

The consistency requirement is that the path integral factorize correctly in all degeneration channels, which is encoded algebraically in the geometric BV master equation~\cite{Zwiebach:1992ie}:
\begin{equation}
    \ud\langle V| + \tfrac{1}{2}\{\langle V|,\langle V|\} + \hbar\,\Lam\langle V| = 0\,,\label{eqn:geometric BV}
\end{equation}
where $\ud$ is the boundary operator on moduli space, $\{{\cdot},{\cdot}\}$ is the antibracket (twist-sewing one puncture from each of two surfaces), and $\Lam$ is the BV operator (twist-sewing two punctures on the same surface). This equation ensures that the decomposition of moduli space is consistent: the Feynman diagrams generated by the propagator, together with the elementary vertices, tile moduli space without gaps or overlaps.

The algebraic content of the BV master equation is that the vertices define a \emph{quantum $L_\infty$ algebra}~\cite{Zwiebach:1992ie}. The $n$-string products $L_{g,n}(\Psi_1,\ldots,\Psi_n)$ are graded-symmetric multilinear maps, and the classical ($g=0$) products satisfy the $L_\infty$ relations. The first of these is the nilpotency $\QB^2 = 0;$ the second is the BRST compatibility of the cubic product:
\begin{equation}
    \QB\, L_{0,2}(\Psi_1,\Psi_2) + L_{0,2}(\QB\Psi_1,\Psi_2) + (-1)^{|\Psi_1|}L_{0,2}(\Psi_1,\QB\Psi_2) = 0\,.\label{eqn:Linfty relation}
\end{equation}
The classical equation of motion following from eq.~\eqref{eqn:CSFT action} is
\begin{equation}
    \QB\Psi + \sum_{n=2}^{\infty}\frac{\ka^{n-1}}{n!}\,L_{0,n}(\underbrace{\Psi,\ldots,\Psi}_{n}) = 0\,,\label{eqn:classical eom}
\end{equation}
and the gauge transformations take the form $\delta\Psi = \QB\Lam + \ka\,L_{0,2}(\Lam,\Psi) + \cdots,$ with Grassmann-odd parameter $\Lam$ of ghost number $1.$ The gauge symmetry is infinitely reducible, necessitating the BV formalism for a well-defined path integral~\cite{Zwiebach:1992ie}.

The construction as presented is abstract: it applies to \emph{any} moduli space decomposition that satisfies eq.~\eqref{eqn:geometric BV}. To obtain a concrete theory, one must specify how the vertex regions and Feynman regions are carved out of $\cM_{g,n}.$ In Zwiebach's original work~\cite{Zwiebach:1992ie}, this was achieved using \emph{minimal-area metrics}. While conceptually clean, minimal-area metrics proved extremely difficult to construct explicitly. The four-string and five string vertices were studied numerically by Moeller~\cite{Moeller:2004yy,Moeller:2006cw}, but extending this program to higher vertices and genera remained difficult.\footnote{For progress along this direction, see for example \cite{Erbin:2022rgx,Firat:2023sld}.} Despite decades of effort and important structural results by Headrick and Zwiebach~\cite{Headrick:2018dlw}, no general explicit solution to the minimal-area problem was found.

A qualitatively new approach was developed by Costello and Zwiebach~\cite{Costello:2019fuh}, building on the foundational work of Faroogh Moosavian and Pius~\cite{Moosavian:2017qsp,Moosavian:2017sev}, using \emph{hyperbolic geometry} rather than minimal-area metrics as the organizing principle. The resulting construction is computationally tractable and forms the foundation for everything that follows.

\subsection{The hyperbolic three-string vertex}\label{sec:review-vertex}

The central idea of Costello and Zwiebach~\cite{Costello:2019fuh} is to replace the minimal-area condition with a constraint formulated entirely in terms of hyperbolic geometry. One requires that each bordered Riemann surface carry a metric of constant curvature $K=-1$ with geodesic borders, and that the systole --- the length of the shortest nontrivial closed geodesic --- be at least $L,$ where $L$ is a free parameter satisfying $0 < L \leq \Lstar$ with
\begin{equation}
    \Lstar =2\,\mathrm{arcsinh}(1) = 1.76274717\ldots\label{eqn:Lstar}
\end{equation}
being the \emph{systolic threshold}. The maximal value $L = \Lstar$ is distinguished: it is the largest value for which the resulting vertex regions tile moduli space consistently, i.e., for which the geometric BV master equation~\eqref{eqn:geometric BV} is satisfied~\cite{Costello:2019fuh}. The hyperbolic approach has the decisive advantage that the moduli space of hyperbolic surfaces is extremely well studied, and the local coordinates around punctures can be constructed explicitly from the uniformization data.

The cubic vertex $\langle V_{0,3}(L_1,L_2,L_3)|$ is the seed datum of the entire construction. In the hyperbolic framework, it corresponds to a unique surface --- the three-holed sphere (pair of pants) with geodesic borders of prescribed lengths $L_i = 2\pi\lam_i$ --- and the moduli space $\cM_{0,3}$ is a single point. The crucial question is how to extract the local coordinates around the punctures, which are needed to evaluate CFT correlators on this surface.

This question was answered by F{\i}rat~\cite{Firat:2021ukc}, who obtained the first explicit, closed-form expression for the local coordinates of the hyperbolic three-string vertex. The construction is based on a classical result in hyperbolic geometry: any Riemann surface with constant negative curvature is uniformized by a Fuchsian group, and the hyperbolic metric can be recovered from the solutions of a Fuchsian ordinary differential equation. For the three-holed sphere with punctures fixed at $z = 0,\,1,\,\infty$ by $\mathrm{PSL}(2,\C),$ this equation takes the form
\begin{equation}
    \psi''(z) + \tfrac{1}{2}\,T_\phi(z)\,\psi(z) = 0\,,\label{eqn:Fuchsian}
\end{equation}
where $T_\phi(z)$ is the holomorphic stress tensor of the hyperbolic metric. For three punctures, there are no unfixed accessory parameters, and $T_\phi(z)$ is determined entirely by the conformal weights $\Delta_j = (1+\lam_j^2)/2$ at each puncture. The Fuchsian equation~\eqref{eqn:Fuchsian} thus reduces to the Gauss hypergeometric equation. Its two linearly independent solutions $\psi_\pm(z)$ are given explicitly in terms of ${}_2F_1$ functions, and the ratio $A(z) = \psi_+(z)/\psi_-(z)$ recovers the stress tensor through the Schwarzian derivative: $T_\phi(z) = \{A,z\}.$

The hyperbolic pair of pants has geodesic borders, which are closed curves of constant geodesic curvature. To construct local coordinates suitable for string field theory, one grafts a flat semi-infinite cylinder of circumference $L_j$ onto each geodesic boundary. The local coordinate $w_j$ around puncture $j$ maps the interior of the unit disk in the $w_j$-plane to a neighborhood of the puncture on the grafted surface. The result, obtained by F{\i}rat (eq.~(1.3) of~\cite{Firat:2021ukc}), is
\begin{equation}
    w_j(z) = \frac{1}{N_j}\exp\!\left(\frac{v(\lam_j,\lam_i,\lam_k)}{\lam_j}\right) z\,(1-z)^{-\lam_i/\lam_j}\!\left[\frac{{}_2F_1(a_+,b_+;1+\ui\lam_j;z)}{{}_2F_1(a_-,b_-;1-\ui\lam_j;z)}\right]^{1/(\ui\lam_j)}\!,\label{eqn:local coordinate}
\end{equation}
where $a_{\pm} = \tfrac{1}{2}(1\pm\ui\lam_j \mp\ui\lam_i\pm\ui\lam_k),$ $b_{\pm} = \tfrac{1}{2}(1\pm\ui\lam_j\mp\ui\lam_i\mp\ui\lam_k),$ and $N_j$ is a normalization factor. The function $v(\lam_1,\lam_2,\lam_3)$ is real and encodes the compatibility of the $\mathrm{SL}(2,\R)$ monodromies around all three punctures simultaneously. It is determined by the condition that the monodromy matrices of the Fuchsian equation lie in $\mathrm{SL}(2,\R)$ (not merely $\mathrm{SL}(2,\C)$). Its explicit form is (eq.~(3.21) of~\cite{Firat:2021ukc})
\begin{equation}
    \ue^{2\ui v(\lam_1,\lam_2,\lam_3)} = \frac{\Ga(-\ui\lam_1)^2}{\Ga(\ui\lam_1)^2}\prod_{s_1,s_2=\pm}\gamma\!\left(\frac{1+s_1\ui\lam_1+s_2\ui\lam_2+\ui\lam_3}{2}\right),\label{eqn:v function}
\end{equation}
where $\ga(x) = \Ga(x)/\Ga(1-x).$ Since $\ue^{2\ui v}$ is a pure phase, eq.~\eqref{eqn:v function} determines $v$ only modulo~$\pi.$

The key geometric datum entering all vertex computations is the \emph{mapping radius} $\rho_j$ at puncture $j,$ defined by $\rho_j = |dz/dw_j|_{w_j=0}.$ It measures the ``size'' of the local coordinate disk as seen from the global uniformization. From eq.~\eqref{eqn:local coordinate}, the mapping radius takes the form (eq.~(3.34) of~\cite{Firat:2021ukc}):
\begin{equation}
    \rho_j = N_j\exp\!\left(-\frac{v(\lam_j,\lam_i,\lam_k)}{\lam_j}\right)\,,\label{eqn:mapping radius}
\end{equation}
where $N_j = \exp[\pi(\tilde{l}_j+\tfrac{1}{2})/\lam_j]$ and $\tilde{l}_j$ is an integer determined by the uniformization geometry; $\tilde{l}_j$ and the branch of $v$ are chosen jointly so that $\rho_j$ is the correct mapping radius.\footnote{For equal border lengths $\lam_1=\lam_2=\lam_3=\lam,$ the integer is $\tilde{l}=-1$ for all $\lam > 0,$ giving $N = \exp(-\pi/(2\lam))$ and $\rho = \exp[-\pi/(2\lam) - v(\lam,\lam,\lam)/\lam].$ In the numerical implementation, we absorb the $\tilde{l}$-dependence into the branch choice of $v$ and use $\tilde{l}=-1$ uniformly; see Appendix~\ref{app:numerics}.} The vertex possesses $S_3$ permutation symmetry acting simultaneously on the puncture labels and the $\lam_i,$ and all Taylor coefficients in the expansion of the local coordinates can be chosen real~\cite{Firat:2021ukc}.

For a primary state $\ket{\cO}$ of conformal weight $(h,\bar{h})$ inserted at puncture $j,$ the local coordinate transformation contributes a factor of $\rho_j^{2h}$ to the vertex (with $h = \bar{h}$ for the states we consider). The full vertex evaluation on any set of external states is then the product of these conformal factors with the CFT correlator on the three-punctured sphere. The fact that $\rho_j$ is a known, explicitly computable function of the border lengths is what makes the hyperbolic cubic vertex a practical starting point for the entire theory. Throughout this paper we use the convention $\alpha'=2,$ so that the zero-momentum tachyon has conformal weight $h=\bar{h} = -1$ and sits at mass-squared $m^2 = -4/\alpha' = -2.$

Before the work of F{\i}rat~\cite{Firat:2021ukc,Firat:2023sld,Firat:2023gfn}, no such non-trivial expression for the closed string cubic vertex existed. The minimal-area vertices of~\cite{Zwiebach:1992ie} could in principle be obtained from a complicated variational problem, and the polyhedral approach of Moeller~\cite{Moeller:2004yy} yielded numerical results at $V_{0,4},$ but none of these gave a closed-form local coordinate in terms of standard special functions. It is this new level of explicitness --- mapping radii expressed in terms of $\Ga$-functions and hypergeometric data --- that makes the program developed in the remainder of this paper feasible.

\subsection{The Fenchel--Nielsen b-ghost $\BG$}\label{sec:review-bghost}

The Fenchel--Nielsen b-ghost is an operator that arises naturally from the hyperbolic geometry of the moduli space. To understand its origin, recall that any bordered hyperbolic surface admits a pants decomposition, and the Fenchel--Nielsen coordinates $(l_i,\tau_i)$ --- lengths and twists along the internal geodesics (``seams'') --- provide global coordinates on the cover of moduli space~\cite{Wolpert:2010wn}. The Weil--Petersson symplectic form in these coordinates takes the simple form $\omega_{\mathrm{WP}} = \sum_i \ud l_i \wedge \ud\tau_i,$ so the twist $\tau_i$ is the coordinate canonically conjugate to the length $l_i.$

In string field theory, integrating over moduli requires a b-ghost insertion to absorb the antighost zero mode associated with each modulus~\cite{Zwiebach:1992ie}. The length of a seam is one such modulus, and the corresponding insertion is the contour integral of the b-ghost field along the vector field that generates the Fenchel--Nielsen. Concretely, if $v_{\mathrm{FN}}$ denotes the Beltrami differential corresponding to an infinitesimal change of the seam at the $j$-th geodesic boundary of a pair of pants, then the Fenchel--Nielsen b-ghost is defined as~\cite{Firat:2024ajp,Costello:2019fuh}
\begin{equation}
    \BG_j = \frac{1}{2\pi\ui}\oint \ud z\;b(z)\,v_{\mathrm{FN}}^z(z) + \frac{1}{2\pi\ui}\oint \ud\bar{z}\;\bar{b}(\bar{z})\,v_{\mathrm{FN}}^{\bar{z}}(\bar{z})\,,\label{eqn:bghost definition}
\end{equation}
where the contour encircles the puncture associated with the deformed border. This definition is intrinsic to the geometry of the pair of pants: $v_{\mathrm{FN}}$ is determined by the hyperbolic metric and the pants decomposition, and $\BG_j$ depends on all three border lengths $(L_j,L_i,L_k).$

The vector field $v_{\mathrm{FN}}$ is most easily computed from the conformal map~\eqref{eqn:local coordinate}. Expanding $\BG_j$ in modes of the $b$-ghost, one obtains~\cite{Firat:2024ajp}
\begin{equation}
    \BG_j = \beta_0(L_j,L_i,L_k)\,(b_0+\bar{b}_0) + \beta_1(L_j,L_i,L_k)\,(b_1+\bar{b}_1) + \cdots\,,\label{eqn:bghost expansion}
\end{equation}
where $b_0+\bar{b}_0 = b_0^+$ in the Zwiebach convention (note the asymmetric normalization: $b_0^+ = b_0+\bar{b}_0$ carries no factor of $\tfrac{1}{2},$ while $c_0^+ = \tfrac{1}{2}(c_0+\bar{c}_0)$; the anticommutator $\{b_0^+,c_0^+\}=1$ then holds with unit coefficient). The leading coefficient is
\begin{equation}
    \beta_0 = \mathrm{BG}(L_j,L_i,L_k) = \frac{1}{2\pi\rho_j}\frac{\ud\rho_j}{\ud\lam_j}\,,\label{eqn:BG leading}
\end{equation}
which is the logarithmic derivative of the mapping radius with respect to the reduced border length $\lam_j = L_j/(2\pi),$ divided by $2\pi.$ When the deformed border needs to be distinguished from the spectator lengths, we write $\BG(\ell;L_i,L_k)$ with a semicolon; when only a single argument appears, as in $\BG(\ell),$ the spectator lengths are determined by the vertex subscripts of the term in which it occurs (see eq.~\eqref{eqn:eq69 gauge fixed}).

The operator $\BG$ plays two conceptually distinct roles in the formulation. First, it provides the natural \emph{gauge-fixing operator} $\beta$ for the equation of motion: the relation $\{\QB,\beta(L)\} = -P,$ where $P$ is the projector onto the complement of the BRST cohomology, holds when $\beta$ is evaluated at the threshold length $L.$ Our gauge choice throughout this paper is $\beta = \BG.$ A different choice (for example, Siegel gauge, $\beta = b_0^+$) would modify the $A$- and $B$-terms of the equation of motion but leave the $C$-term unchanged. Second, $\BG$ appears as an \emph{intrinsic seam insertion} in the topological recursion: the $B$- and $C$-products of \S\ref{sec:review-recursion} below carry $\BG$ insertions at the integrated punctures, arising from the negative-length convention of eq.~(4.13) of~\cite{Firat:2024ajp}. These insertions are built into the recursion and are independent of the gauge choice.

At the level of the tachyon state $\ket{T_3} = c_0^+\,c_1\tilde{c}_1\ket{0}$ (ghost number $3$), only the leading mode contributes:
\begin{equation}
    \BG\ket{T_3} = \beta_0\,(b_0+\bar{b}_0)\,c_0^+\,c_1\tilde{c}_1\ket{0} = \beta_0\ket{T_2}\,,\label{eqn:BG on tachyon}
\end{equation}
where $\ket{T_2} = c_1\tilde{c}_1\ket{0}$ has ghost number $2$ and $\{b_0^+,c_0^+\}=1$ has been used. Thus, at the tachyon level, $\BG$ reduces to multiplication by the scalar $\mathrm{BG}(L_j,L_i,L_k).$ At higher Fock space levels, the subleading coefficients $\beta_1,\beta_2,\ldots$ contribute nontrivially, and $\BG$ acts as a genuine operator rather than a scalar.

\subsection{Topological recursion for hyperbolic string vertices}\label{sec:review-recursion}

The defining property of the hyperbolic construction that makes the present work possible is a topological recursion: every higher-order string vertex $\langle V_{g,n}|$ with $2g-2+n > 1$ is determined iteratively from the cubic vertex $\langle V_{0,3}|$ alone. This recursion, established by F{\i}rat and Valdes-Meller~\cite{Firat:2024ajp}, is a string-theoretic analog of the celebrated recursion of Mirzakhani~\cite{Mirzakhani:2006fta} for Weil--Petersson volumes, itself an instance of the Eynard--Orantin topological recursion~\cite{EynardOrantin:2007}.\footnote{See also \cite{andersen2017geometric,Ishibashi:2022qcz}.} We now review its derivation and structure.

The starting point is Mirzakhani's identity for hyperbolic surfaces. Consider a bordered hyperbolic surface $\Sigma$ of genus $g$ with $n$ geodesic borders of lengths $L_1,\ldots,L_n.$ Mirzakhani showed that the Weil--Petersson volume $V_{g,n}^{\mathrm{WP}}(L_1,\ldots,L_n)$ of the moduli space of such surfaces satisfies a recursive identity driven by simple closed geodesics. The idea is to consider all simple closed geodesics on $\Sigma$ that are disjoint from the first border (the one of length $L_1$). There are two types of contributions, one that cuts along one geodesic (an R-type contribution) and the other that cuts along two nearby geodesics (a D-type contribution). Integrating over all possible lengths and twist angles of $\ga,$ weighted by the Weil--Petersson measure, one obtains
\begin{equation}
    L_1\,\frac{\partial}{\partial L_1}V_{g,n}^{\mathrm{WP}}(L_i) = (\text{$R$- and $D$-integrals of lower $V^{\mathrm{WP}}$})\,,\label{eqn:Mirzakhani identity}
\end{equation}
where the right-hand side involves volumes at lower Euler characteristic. The integration kernels are the \emph{Mirzakhani kernels}~\cite{Mirzakhani:2006fta}:
\begin{align}
    D_{L_1,L_2,L_3} &= 2\log\!\left[\frac{e^{L_1/2}+e^{(L_2+L_3)/2}}{e^{-L_1/2}+e^{(L_2+L_3)/2}}\right]\,,\label{eqn:D kernel}\\[6pt]
    R_{L_1,L_2,L_3} &= L_1 - \log\!\left[\frac{\cosh(L_2/2)+\cosh\!\big((L_1+L_3)/2\big)}{\cosh(L_2/2)+\cosh\!\big((L_1-L_3)/2\big)}\right]\,.\label{eqn:R kernel}
\end{align}
Here $L_1$ is the border being deformed and $L_2,L_3$ are spectator lengths. The $D$-kernel is symmetric in $L_2$ and $L_3,$ while the $R$-kernel is not. 

To pass from volume recursion to string vertex recursion, F{\i}rat and Valdes-Meller~\cite{Firat:2024ajp} make two key observations. First, the hyperbolic string vertices involve not the full moduli space $\cM_{g,n}$ but only the \emph{systolic subset} --- the region where every simple-non-contractable closed geodesic has length at least $L.$ Surfaces with a closed geodesic shorter than $L$ belong to the Feynman region and are generated by the propagator. This restriction is implemented by introducing \emph{twisted Mirzakhani kernels}, which are the standard kernels with the Feynman-region contributions subtracted:
\begin{align}
    \eR_{L_1,L_2,L_3} &= R_{L_1,L_2,L_3} - L_1\,\theta(L - L_3)\,,\label{eqn:eR}\\[4pt]
    \eD_{L_1,L_2,L_3} &= D_{L_1,L_2,L_3} - R_{L_1,L_2,L_3}\,\theta(L-L_2) - R_{L_1,L_3,L_2}\,\theta(L-L_3) + L_1\,\theta(L-L_2)\,\theta(L-L_3)\,,\label{eqn:eD}
\end{align}
where $\theta$ is the Heaviside step function and $L$ is the threshold length. The twisted kernels reduce to the standard ones as $L\to 0,$ in which case the systolic subsets exhaust the full moduli space. At $L = \Lstar,$ they restrict the integration to the maximal vertex region consistent with the BV master equation.

Second, the twisted kernels enter the vertex recursion not as numerical weights on volumes but as scalar multiplicative factors on the cubic vertex itself. This is the crucial \emph{factorization property}: the \emph{string kernels}~\cite{Firat:2024ajp} are defined as
\begin{equation}
    \langle R(L_1,L_2,L_3)| = \eR_{\abs{L_1},\abs{L_2},\abs{L_3}}\;\langle V_{0,3}(L_1,L_2,L_3)|\,,
\end{equation}
\begin{equation}
    \langle D(L_1,L_2,L_3)| = \eD_{\abs{L_1},\abs{L_2},\abs{L_3}}\;\langle V_{0,3}(L_1,L_2,L_3)|\,,\label{eqn:string kernels}
\end{equation}
and the recursion relation for hyperbolic string vertices (eq.~(4.16) of~\cite{Firat:2024ajp}) then reads
\begin{align}
    \abs{L_1}\,\langle V_{g,n}(L_i)| &= \sum_{i=2}^{n}\int_{-\infty}^{\infty}\!\ud\ell\;\big(\langle R(L_1,L_i,\ell)|\otimes\langle V_{g,n-1}(-\ell,\widehat{L}_i)|\big)\,\ket{\omega^{-1}}\nn\\[4pt]
    &\quad + \frac{1}{2}\int_{-\infty}^{\infty}\!\ud\ell_1\int_{-\infty}^{\infty}\!\ud\ell_2\;\bigg(\langle D(L_1,\ell_1,\ell_2)|\otimes\bigg[\langle V_{g-1,n+1}(-\ell_1,-\ell_2,L)|\nn\\[2pt]
    &\qquad\qquad + \sum_{\mathrm{stable}}\langle V_{g_1,n_1}(-\ell_1,L_{\cS_1})|\otimes\langle V_{g_2,n_2}(-\ell_2,L_{\cS_2})|\bigg]\bigg)\,\ket{\omega^{-1}}_1\ket{\omega^{-1}}_2\,,\label{eqn:topological recursion}
\end{align}
where $\widehat{L}_i$ denotes the set $\{L_2,\ldots,L_n\}\setminus\{L_i\},$ $\ket{\omega^{-1}} $ is the Poisson bivector implementing the twist-sewing operation~\cite{Zwiebach:1992ie}, and the sum runs over stable splittings $g_1+g_2 = g,$ $n_1+n_2 = n+1.$ A negative border length $L_i < 0$ signals that the $\BG$ operator is inserted at puncture $i,$ with the physical border length being $|L_i|;$ this is the convention of eq.~(4.13) of~\cite{Firat:2024ajp}.

The physical content of eq.~\eqref{eqn:topological recursion} is as follows. The left-hand side is a higher-order string vertex, encoding an elementary $n$-string interaction at genus $g.$ The right-hand side constructs this vertex by considering all possible ways to cut the surface along an internal geodesic. In each case, the cut produces two smaller surfaces, each carrying a lower-order vertex, sewn together through the Poisson bivector $\ket{\omega^{-1}}.$ Because the string kernels $\langle R|$ and $\langle D|$ factorize through the cubic vertex by eq.~\eqref{eqn:string kernels}, the entire infinite tower of vertices is determined recursively from $\langle V_{0,3}|$ alone.

This is a remarkable structural result. The same topological recursion that Mirzakhani used to compute Weil--Petersson volumes of moduli space~\cite{Mirzakhani:2006fta}---subsequently reformulated as an instance of the Eynard--Orantin recursion~\cite{EynardOrantin:2007}---now governs the full hierarchy of closed string field theory interactions. The infinite tower of nonpolynomial vertices, which was the defining difficulty of closed string field theory, is \emph{computable} from a single input --- the cubic vertex --- together with the Mirzakhani kernels.

\subsection{The quadratic integral equation}\label{sec:review-eq69}

The topological recursion~\eqref{eqn:topological recursion} has a far-reaching consequence for the classical equation of motion~\eqref{eqn:classical eom}. Since the recursion generates all higher vertices from the cubic one, the infinite series in eq.~\eqref{eqn:classical eom} is not independent data: it is completely determined by $\langle V_{0,3}|.$ F{\i}rat and Valdes-Meller~\cite{Firat:2024ajp} exploit this to encode the full equation of motion as a single self-consistency condition.

The construction proceeds in two steps. First, the topological recursion is repackaged as a differential constraint (eq.~(4.34) of~\cite{Firat:2024ajp}) on a generating function $W[\Psi(\ell)]$ of string vertices. Second, this constraint is evaluated at a classical solution $\Psi_*$ of the equation of motion. One introduces the border-length-dependent string field $\Phi(L_1),$ defined as a generating function of the solution evaluated at generalized vertices (eq.~(6.2) of~\cite{Firat:2024ajp}):
\begin{equation}
    \Phi(L_1) = \sum_{n=2}^{\infty}\frac{\ka^{n-1}}{n!}\,L_{0,n}(\underbrace{\Psi_*,\ldots,\Psi_*}_{n};\,L_1)\,,\label{eqn:Phi generating}
\end{equation}
where $L_{0,n}(\,\cdot\,;L_1)$ denotes the generalized $L_\infty$ product in which the first border has length $L_1$ rather than the threshold length $L.$ At $L_1 = L,$ one recovers $\Phi(L) = -\QB\Psi_*$ by the equation of motion. Using the gauge-fixing condition $\beta\,\Phi(L) = \Psi_*$ (eq.~(6.8) of~\cite{Firat:2024ajp}), where $\beta$ is a gauge-fixing operator satisfying $\{\QB,\beta(L)\} = -P,$ the differential constraint reduces to the \emph{quadratic integral equation} for the string field --- eq.~(6.9) of~\cite{Firat:2024ajp}:
\begin{align}
    \Phi(L_1) &= \frac{\ka}{2}\,A_{L_1,L,L}\!\left(\beta\,\Phi(L),\;\beta\,\Phi(L)\right)\nn\\[4pt]
    &\quad + \ka\int_{-\infty}^{\infty}\!\ud\ell\; B_{L_1,L,\ell}\!\left(\beta\,\Phi(L),\;\Phi(\ell)\right)\nn\\[4pt]
    &\quad + \frac{\ka}{2}\int_{-\infty}^{\infty}\!\ud\ell_1\int_{-\infty}^{\infty}\!\ud\ell_2\; C_{L_1,\ell_1,\ell_2}\!\left(\Phi(\ell_1),\;\Phi(\ell_2)\right)\,,\label{eqn:eq69}
\end{align}
where $\beta$ is the gauge-fixing operator (our choice: $\beta = \BG$).\footnote{In eq.~\eqref{eqn:eq69}, $\beta$ denotes a generic gauge-fixing operator. In the gauge-fixed equation~\eqref{eqn:eq69 gauge fixed} and all subsequent formulas, we set $\beta = \BG$ and write $\BG$ explicitly to emphasize its operator nature. In the multi-level source formulas of \S\ref{sec:seam-algebra}, $\beta$ is reused as a Fock-space summation index; the context (operator vs.\ index) always disambiguates.} The string field $\Phi(L_1)$ is a state in the infinite-dimensional closed-string Fock space $\Fock.$ Let $\{\ket{\alpha}\}$ denote a basis of ghost-number-$3$ states satisfying $b_0^-\ket{\alpha} = 0$ and $L_0^-\ket{\alpha} = 0.$ The general expansion is
\begin{equation}
    \Phi(L_1) = \sum_\alpha g_\alpha(L_1)\,\ket{\alpha}\,,\label{eqn:Phi expansion}
\end{equation}
where $L_1\in [0,\infty)$ is the border length, the sum runs over all basis states, and the component functions $g_\alpha(L_1)$ encode the border-length dependence at each Fock space level. For example, the first few components are
\begin{equation}
    \Phi(L_1) = g_T(L_1)\,c_0^+c_1\bar{c}_1\ket{0} + \phi_{\mu\nu}(L_1)\,c_0^+\alpha^\mu_{-1}\bar{\alpha}^\nu_{-1}c_1\bar{c}_1\ket{0} + g_{\mathrm{pg}}(L_1)\,c_0^+c_{-1}\bar{c}_{-1}\ket{0} + \cdots\,,\label{eqn:Phi expansion example}
\end{equation}
where the tachyon state $\ket{T} = c_0^+c_1\bar{c}_1\ket{0}$ has ghost number $3$ and similar states happen to contain a factor of $c_0^+,$ but no such factorization is assumed in general: the basis $\{\ket{\alpha}\}$ may include ghost-number-$3$ states with different ghost structures at higher oscillator levels.

The three terms in eq.~\eqref{eqn:eq69} have a direct topological interpretation in terms of the pants decomposition. The $A$-term (no integration) evaluates the cubic vertex with both arguments at the fixed threshold length $L.$ It corresponds to the seed vertex $V_{0,3}$ with no internal seams. The $B$-term (single integral) opens one internal seam: one argument is evaluated at $L,$ the other is integrated over all border lengths, weighted by the twisted $R$-kernel. It encodes the contribution of a single separating geodesic. The $C$-term (double integral) opens two internal seams, with both arguments integrated and weighted by the twisted $D$-kernel; it encodes non-separating geodesics.

The 2-products $A,$ $B,$ and $C$ are defined through the cubic vertex and the Poisson bivector~\cite{Firat:2024ajp}:
\begin{equation}
    A_{L_1,L_2,L_3}(\Psi_1,\Psi_2) = \langle V_{0,3}(L_1,L_2,L_3)|\,\omega^{-1}\,|\Psi_1\rangle|\Psi_2\rangle\,,\label{eqn:A product}
\end{equation}
and the $B$- and $C$-products factorize as
\begin{equation}
    B_{L_1,L_2,L_3} = \eR_{L_1,L_2,L_3}\;A_{L_1,L_2,L_3}\,,\qquad C_{L_1,L_2,L_3} = \eD_{L_1,L_2,L_3}\;A_{L_1,L_2,L_3}\,,\label{eqn:BCA factorization}
\end{equation}
with the twisted kernels acting as scalar multiplicative weights. This factorization, inherited from eq.~\eqref{eqn:string kernels}, is the reason the full nonpolynomial equation of motion collapses to a quadratic integral equation involving only the cubic vertex data.

After absorbing $\ka$ by the field redefinition $\Phi \to (4/\ka)\,\Phi$ (following~\cite{Firat:2024ajp}), the combinatorial coefficients become
\begin{equation}
    C_A = 2\,,\qquad C_B = 4\,,\qquad C_C = 2\,.\label{eqn:combinatorial coefficients}
\end{equation}
Using the negative-length convention (eq.~(4.13) of~\cite{Firat:2024ajp}), in which $\BG$ acts on the string field at each internal geodesic, the integration domain folds from $(-\infty,\infty)$ to $[0,\infty).$ Under the gauge choice $\beta = \BG,$ for general threshold length $L,$ eq.~\eqref{eqn:eq69} then takes the form
\begin{align}
    \Phi(L_1) &= C_A\,A_{L_1,L,L}\!\left(\BG\,\Phi(L),\;\BG\,\Phi(L)\right)\nn\\[4pt]
    &\quad + C_B\int_0^{\infty}\!\ud\ell\;\eR(L_1,L,\ell)\;A_{L_1,L,\ell}\!\left(\BG\,\Phi(L),\;\BG\,\Phi(\ell)\right)\nn\\[4pt]
    &\quad + C_C\int_0^{\infty}\!\ud\ell_1\int_0^{\infty}\!\ud\ell_2\;\eD(L_1,\ell_1,\ell_2)\;A_{L_1,\ell_1,\ell_2}\!\left(\BG\,\Phi(\ell_1),\;\BG\,\Phi(\ell_2)\right)\,,\label{eqn:eq69 gauge fixed}
\end{align}
where the argument of $\BG$ denotes the border at which the operator acts; its spectator lengths are fixed by the vertex subscripts (e.g., in the $B$-term, $\BG(\ell)$ is shorthand for $\BG(\ell;L,L_1)$). This is the starting point for the seam-graded expansion.

\subsubsection*{Component notation and $\BG$-dressed components}

To project eq.~\eqref{eqn:eq69 gauge fixed} onto Fock space components, we need the matrix elements of the operator $\BG$ between basis states. Let $\{\ket{\alpha}\}$ be the ghost-number-$3$ basis introduced in eq.~\eqref{eqn:Phi expansion}, and let $\{\ket{\beta}\}$ be the corresponding ghost-number-$2$ basis (states satisfying $b_0^-\ket{\beta}=0,$ $L_0^-\ket{\beta}=0,$ and $\mathrm{gh}(\beta)=2$). Since $\BG$ lowers ghost number by $1,$ the state $\BG\ket{\alpha}$ has ghost number $2.$ The overlap with a ghost-number-$2$ bra requires the $c_0^-$ insertion that makes the closed-string inner product non-degenerate on the $(b_0^- = 0,\,L_0^- = 0)$ subspace. We define the $\BG$-matrix elements as
\begin{equation}
    B_{\beta\alpha}(L) = \bra{\beta}\,c_0^-\,\BG(L)\,\ket{\alpha}\,,\label{eqn:BG matrix}
\end{equation}
where $L$ stands for the full dependence on the border at which $\BG$ acts and the spectator lengths, and the resolution of the identity on the constraint surface is $\mathbf{1} = \sum_\beta\ket{\beta}\bra{\beta}\,c_0^-.$ The $\BG$-dressed components of $\Phi$ are then defined as
\begin{equation}
    [\BG(L)\,\Phi(L)]_\beta \;=\; \sum_\alpha B_{\beta\alpha}(L)\;g_\alpha(L)\,,\label{eqn:BG dressed}
\end{equation}
which is a scalar (a number, not an operator) for each index $\beta$ and each value of $L.$ The bare vertex components $V^{\alpha,\beta,\ga}(L_1,L_2,L_3)$ are defined by projecting $A_{L_1,L_2,L_3}(\ket{\beta},\ket{\ga})$ onto the ghost-number-$3$ basis: $\alpha$ ranges over ghost-number-$3$ states, while $\beta$ and $\ga$ range over ghost-number-$2$ states. The component form of $A(\BG\Phi,\BG\Phi)$ therefore involves the $\BG$-dressed components:
\begin{equation}
    \big[A_{L_1,L_2,L_3}(\BG\,\Phi(L_2),\;\BG\,\Phi(L_3))\big]_\alpha = \sum_{\beta,\ga}V^{\alpha,\beta,\ga}(L_1,L_2,L_3)\;[\BG\,\Phi(L_2)]_\beta\;[\BG\,\Phi(L_3)]_\ga\,.\label{eqn:component vertex}
\end{equation}
At the tachyon level, $\BG$ reduces to the scalar $\mathrm{BG}(L)$ by eq.~\eqref{eqn:BG on tachyon}, so $B_{T,T}(L) = \mathrm{BG}(L)$ and all off-diagonal elements vanish. More generally, between level-matched states the $B$-matrix receives contributions only from the leading mode $\beta_0\,b_0^+$: the matrix element $\bra{\beta}c_0^-\,b_n\ket{\alpha}$ vanishes for $n\neq 0$ by the level-matching constraint $L_0^-\ket{\alpha} = 0,$ since $[L_0^-,\,c_0^-\,b_n] = -n\,c_0^-\,b_n.$ Consequently, $B_{\beta\alpha}(L) = \mathrm{BG}(L)\,\bra{\beta}c_0^-\,b_0^+\ket{\alpha}$ is block-diagonal in oscillator level, and the $\BG$-dressed components at each level are simply $\mathrm{BG}(L)$ times the $b_0^+$ matrix elements.

Throughout the remainder of this paper, when component formulas appear, the notation $[\BG(L)\,\Phi(L)]_\beta$ always denotes the scalar quantity defined by eq.~\eqref{eqn:BG dressed}. In the tachyon-only projection, where $\BG$ acts as a scalar, this reduces to $\mathrm{BG}(L)\cdot g_T(L)$ and the matrix structure is trivial.

In the general vertex, the string field enters with arbitrary spacetime derivatives $\partial_\mu\Phi,\,\partial_\mu\partial_\nu\Phi,\,\ldots$ arising from the oscillator mode expansion and the structure of the vertex operators. Throughout this paper, we restrict to the sector of zero-momentum Lorentz-scalar states, following Yang and Zwiebach~\cite{Yang:2005rx}. In this sector, all spacetime derivative terms are absent and eq.~\eqref{eqn:component vertex} applies directly. Lorentz symmetry ensures that this sector is closed under the equations of motion. This is the natural first case to study: string field theory is formulated perturbatively around a given background, and even a purely scalar condensate could fundamentally restructure --- or annihilate --- spacetime. Whether the physical tachyon vacuum lies within the scalar sector is an open question. There is no a priori reason to expect the vacuum to be translation-invariant: the tachyon may nucleate locally, form bubble configurations, or the vacuum itself may obscure the spacetime picture entirely. The physical vacuum may well require non-scalar condensation. Studying solutions beyond the scalar sector is an important direction for future work.

\section{The seam-graded algebra}\label{sec:seam-algebra}

\subsection{The seam grading}\label{sec:seam-grade-def}

Equation~\eqref{eqn:eq69 gauge fixed} has three terms, each with a different number of border-length integrations: the A-term has none, the B-term has one, and the C-term has two. A natural question is whether this structure can be used to organize the solution iteratively. The answer is yes, and the organizing principle is topological: at each stage of the iteration, one can count how many internal geodesics --- seams --- have been integrated over.

We make this precise as follows. Let $\cH = L^2\!\left([0,\infty);\Fock\right)$ be the space of square-integrable Fock-space-valued functions of the border length, and write the string field as a formal expansion
\begin{equation}
    \Phi(L_1) = \sum_{n=0}^{\infty}\Phi_n(L_1)\,,\label{eqn:seam expansion}
\end{equation}
where $\Phi_n$ has \emph{seam grade} $n.$ The integer $n$ counts the number of internal geodesics that were integrated over in the pants decomposition to produce the contribution $\Phi_n.$ Equivalently, $n$ counts the number of pairs of pants beyond the initial three-punctured sphere:
\begin{center}
\begin{tabular}{c|c|c|c|c}
Grade $n$ & Surface & \# pants & \# seams & \# ext.\ borders \\
\hline
$0$ & $V_{0,3}$ & $1$ & $0$ & $3$ \\
$1$ & $V_{0,4}$ & $2$ & $1$ & $4$ \\
$2$ & $V_{0,5}$ & $3$ & $2$ & $5$ \\
$n$ & $V_{0,n+3}$ & $n+1$ & $n$ & $n+3$
\end{tabular}
\end{center}
When $\Phi = \sum_{n\geq 0}\Phi_n$ is substituted into the right-hand side of eq.~\eqref{eqn:eq69 gauge fixed}, each resulting term acquires a definite seam grade by the following counting rules. Consider a term in which the operator (A, B, or C) acts on arguments $\Phi_j$ and $\Phi_k.$ The total grade of the resulting contribution is
\begin{equation}
    \text{grade} = (\text{intrinsic seam cost of the operator}) + j + k\,.\label{eqn:grade counting}
\end{equation}
It remains to determine the intrinsic seam cost of each operator, and to explain why grades add.

The A-term $A_{L_1,L,L}$ evaluates the cubic vertex at three fixed border lengths. No integration is performed, and no new internal geodesic is opened. Its intrinsic seam cost is therefore $0,$ and a term $A(\BG\Phi_j,\BG\Phi_k)$ has total grade $0 + j + k = j+k.$

The B-term $\int_0^\infty\!\ud\ell\;\eR\,A_{L_1,L,\ell}$ integrates one border length $\ell$ over $[0,\infty),$ weighted by the twisted R-kernel $\eR.$ Geometrically, this integration over $\ell$ corresponds to summing over all possible lengths of one new internal geodesic --- a single new seam. The intrinsic seam cost is therefore $1,$ and a term $B(\BG\Phi_j,\BG\Phi_k)$ has total grade $1 + j + k.$

The C-term $\int_0^\infty\!\ud\ell_1\int_0^\infty\!\ud\ell_2\;\eD\,A_{L_1,\ell_1,\ell_2}$ integrates two border lengths simultaneously. Each integration variable corresponds to the length of a distinct internal geodesic: two new seams are opened at once. The intrinsic seam cost is $2,$ and a term $C(\BG\Phi_j,\BG\Phi_k)$ has total grade $2+j+k.$

The additivity of grades in eq.~\eqref{eqn:grade counting} is not an arbitrary convention but a consequence of the topology. The field $\Phi_j$ was assembled by sewing $j+1$ pairs of pants along $j$ seams. When the B-term integrates $\Phi_j$ and $\Phi_k$ together with one new seam, the resulting surface has $j + k + 1$ seams in total. Crucially, no double-counting is possible: the $j$ seams internal to $\Phi_j,$ the $k$ seams internal to $\Phi_k,$ and the new seam opened by the operator correspond to topologically distinct internal geodesics in the pants decomposition. The same reasoning applies to the C-term with two new seams.

To summarize, the three counting rules are:
\begin{equation}
    \text{A-term: } j+k\,,\qquad\text{B-term: } j+k+1\,,\qquad \text{C-term: } j+k+2\,.\label{eqn:counting rules}
\end{equation}

\subsection{The grade-$0$ equation}\label{sec:grade-0}

We now solve the equation grade by grade, beginning with the simplest case. At grade $0,$ every term on the right-hand side of eq.~\eqref{eqn:eq69 gauge fixed} must have total grade $0.$ By the counting rules~\eqref{eqn:counting rules}, a B-term always carries grade $\geq 1$ and a C-term carries grade $\geq 2,$ so neither can contribute. The only possibility is the A-term with both arguments from grade~$0$:
\begin{equation}
    \Phi_0(L_1) = C_A\,A_{L_1,L,L}\!\left(\BG\,\Phi_0(L),\;\BG\,\Phi_0(L)\right)\,.\label{eqn:grade-0 eq}
\end{equation}
This is a purely algebraic equation: no border-length integration is performed, no seams are opened, and in the scalar sector no spacetime derivatives appear. It is the self-consistency condition for the string field evaluated at the cubic vertex $V_{0,3}(L_1,L,L).$ Projecting onto the Fock space components and evaluating at $L_1 = L,$ it reduces to a system of coupled quadratic equations whose structure depends on how many Fock space levels are retained. We now build up the multi-level content of this system in stages, beginning with the tachyon alone and systematically incorporating higher-level states.

Consider first the tachyon sector in isolation. The string field carries a single component $g_T(L_1),$ and $\BG$ acts as the scalar $\mathrm{BG}(L)$ by eq.~\eqref{eqn:BG on tachyon}, so $[\BG\,\Phi_0(L)]_T = \mathrm{BG}\,g_T^{(0)}(L).$ Using eq.~\eqref{eqn:component vertex}, the grade-$0$ equation projects to
\begin{equation}
    g_T^{(0)}(L_1) = C_A\,V_T(L_1,L,L)\;\big[\mathrm{BG}\,g_T^{(0)}(L)\big]^2\,,\label{eqn:grade-0 tachyon}
\end{equation}
where $V_T(L_1,L_2,L_3) = \rho_1^{-2}\rho_2^{-2}\rho_3^{-2}$ is the bare tachyon vertex (the mapping radius $\rho_j$ is raised to the power $-2h = 2$ for the tachyon, with $h=-1$). It is convenient to define the effective tachyon vertex $V_T^{\mathrm{eff}}(L_1,L_2,L_3) = V_T(L_1,L_2,L_3)\,\mathrm{BG}(L_2)\,\mathrm{BG}(L_3),$ which absorbs the $\BG$-dressing at the second and third punctures (one factor per threshold-length slot). Then eq.~\eqref{eqn:grade-0 tachyon} becomes $g_T^{(0)}(L_1) = C_A\,V_T^{\mathrm{eff}}(L_1,L,L)\,[g_T^{(0)}(L)]^2.$ The right-hand side depends on $L_1$ only through the vertex, while the unknown $g_T^{(0)}(L)$ appears as a constant prefactor. Setting $L_1 = L$ reduces the equation to a quadratic self-consistency condition:
\begin{equation}
    g_T^{(0)}(L) = C_A\,V_T^{\mathrm{eff}}(L,L,L)\;\big[g_T^{(0)}(L)\big]^2\,.\label{eqn:grade0 self-consistency}
\end{equation}
The two solutions are $g_T^{(0)}(L) = 0$ (trivial) and the nontrivial root
\begin{equation}
    g_T^{(0)}(L) = \frac{1}{C_A\,V_T^{\mathrm{eff}}(L,L,L)}\,.\label{eqn:grade-0 seed}
\end{equation}
At the systolic point $L = \Lstar,$ the bare tachyon vertex is $V_T(\Lstar,\Lstar,\Lstar) = \rhostar^{-6} \simeq 4.87\times 10^{8}$ and the BG scalar is $\mathrm{BG}(\Lstar) \simeq 1.126,$ giving $V_T^{\mathrm{eff}}(\Lstar,\Lstar,\Lstar) = \mathrm{BG}(\Lstar)^2\,V_T \simeq 6.17\times 10^{8}.$ The nontrivial root is therefore $g_T^{(0)}(\Lstar) \simeq 8.10\times 10^{-10}$ --- a tiny number set by the exponentially large vertex at the systolic threshold. Once this seed value is determined, the full $L_1$-profile follows from eq.~\eqref{eqn:grade-0 tachyon}:
\begin{equation}
    g_T^{(0)}(L_1) = \frac{V_T^{\mathrm{eff}}(L_1,L,L)}{V_T^{\mathrm{eff}}(L,L,L)}\;g_T^{(0)}(L)\,.\label{eqn:grade-0 profile}
\end{equation}
This ratio is an explicit function of $L_1$ built from products of $\Ga$-functions at imaginary arguments, inherited from the hyperbolic mapping radii~\cite{Firat:2023sld}. This tachyon-only equation is the simplest nontrivial seed. We will see in \S\ref{sec:grade-1} that the tachyon-only equation has no nontrivial solution once the B- and C-terms are included at higher grades; the multi-level coupling to higher Fock space states is essential for a consistent solution. Nevertheless, at the A-term level the nontrivial root~\eqref{eqn:grade-0 seed} provides a meaningful starting point.

\subsubsection*{Level-$2$ coupling and selection rules}

Now suppose the string field carries level-$0$ and level-$2$ components simultaneously. At level $2$ in the closed string Fock space, one must enumerate the zero-momentum Lorentz-scalar states satisfying the Siegel gauge condition $b_0^- = 0$ and the level-matching constraint $L_0^- = 0.$ Oscillator level $2$ means $L_0 = \bar{L}_0 = 1$ above the $\mathrm{SL}(2,\C)$ vacuum, so each sector carries exactly one unit of oscillator excitation. The graviton $h_{\mu\nu}$ (symmetric traceless) and the Kalb--Ramond field $B_{\mu\nu}$ (antisymmetric) both carry a free Lorentz index and therefore vanish identically for the Lorentz-singlet tachyon vacuum. The surviving states are three scalars:
\begin{itemize}[nosep]
  \item $D_+$ (dilaton-parity-even ghost state): $\frac{1}{\sqrt{2}}\bigl(c_1 c_{-1} + \bar{c}_1 \bar{c}_{-1}\bigr)\ket{0},$ a superposition of $(G_L,G_R) = (2,0)$ and $(0,2)$ ghost-number sectors, even under worldsheet parity.
  \item $S_1$ (matter dilaton): $c_1 \alpha^\mu_{-1} \bar{c}_1 \bar{\alpha}_{\mu,-1} \ket{0},$ with $(G_L,G_R) = (1,1)$ and a trace contraction over the $D$ matter directions. This is the only level-$2$ state that carries matter oscillator content.
  \item $D$ (ghost dilaton): $\frac{1}{\sqrt{2}}\bigl(c_1 c_{-1} - \bar{c}_1 \bar{c}_{-1}\bigr)\ket{0},$ the parity-odd partner of $D_+,$ again a superposition of $(2,0)$ and $(0,2).$ This is the state whose condensation is required by Yang and Zwiebach's level-truncation analysis of the tachyon vacuum~\cite{Yang:2005rx}.
\end{itemize}
The states listed above are the ghost-number-$2$ basis states $\ket{\beta}.$ At level $2,$ the corresponding ghost-number-$3$ elements of the basis $\{\ket{\alpha}\}$ in eq.~\eqref{eqn:Phi expansion} are simply $c_0^+\ket{\beta};$ however, this factorization is specific to low oscillator levels and does not hold in general at higher oscillator level.
The grade-$0$ equation~\eqref{eqn:grade-0 eq} now becomes a coupled $4$-component quadratic system for the vector $g_\alpha^{(0)}(L)$ with $\alpha \in \{T, D_+, S_1, D\}$:
\begin{equation}
    g_\alpha^{(0)}(L_1) = C_A\sum_{\beta,\ga}V^{\alpha,\beta,\ga}(L_1,L,L)\;[\BG\,\Phi_0(L)]_\beta\;[\BG\,\Phi_0(L)]_\ga\,,\label{eqn:grade-0 multi-level}
\end{equation}
where the sum runs over all retained Fock space states $\beta$ and $\ga,$ $V^{\alpha,\beta,\ga}$ is the bare cubic vertex, and $[\BG\,\Phi_0(L)]_\beta = \sum_{\alpha'}B_{\beta\alpha'}\,g_{\alpha'}^{(0)}(L)$ are the $\BG$-dressed components defined in eq.~\eqref{eqn:BG dressed}. Ghost number conservation at the cubic vertex provides strong selection rules. Three $(1,1)$ insertions produce the required sphere anomaly $(3,3),$ so any triple of $(1,1)$ states is allowed in principle; in particular $V^{T,T,T}$ and $V^{T,S_1,S_1}$ are nonzero. However, the ghost states $D_+$ and $D$ carry $(G_L,G_R)$ sectors $(2,0)$ and $(0,2),$ so a vertex involving two tachyons and one ghost state would give total ghost number $(2,2)+(2,0)=(4,2)$ or $(2,4)$ --- neither equals $(3,3),$ and therefore $V^{T,T,D_+} = V^{T,T,D} = 0.$ Matter correlator selection rules impose a second layer of constraints: $V^{T,T,S_1}=0$ because the tachyon has no matter oscillators for the dilaton's $\alpha_{-1}\bar\alpha_{-1}$ to Wick-contract against, and $V^{S_1,S_1,S_1}=0$ because three $\partial X$ fields cannot be fully Wick-paired.

The nonzero effective vertices at this truncation, with their numerical values at the symmetric point $(\Lstar,\Lstar,\Lstar),$ are listed in table~\ref{tab:level-2-vertices}. The tachyon self-coupling $V^{T,T,T}_{\mathrm{eff}} \simeq 6.17\times 10^{8}$ dominates overwhelmingly because the tachyon vertex carries the conformal factor $\rho^{2}$ (with $h=-1$), which is exponentially large at the systolic threshold. The feedback vertex $V^{T,S_1,S_1} \simeq 2.59 \times 10^{4}$ is the next-largest entry: if the matter dilaton is given a nonzero amplitude, it feeds back into the tachyon equation through this coupling. A smaller but structurally important pair of vertices couples $D_+$ and $D$ to the tachyon equation: $V^{T,D_+,D_+} = -V^{T,D,D} \simeq -7.48\times 10^{2},$ with opposite signs reflecting the parity structure. Crucially, the reverse vertices with two tachyons and one level-$2$ field vanish: $V^{S_1,T,T}=V^{D_+,T,T}=V^{D,T,T}=0.$ Thus a pure tachyon seed does not directly source level-$2$ components at the cubic/grade-$0$ level; level-$2$ fields enter only on branches where they are already nonzero. This inter-level feedback is the mechanism by which higher Fock space states can participate in non-tachyonic grade-$0$ seeds even though the tachyon amplitude is by far the largest single vertex.
\begin{table}[t]
\centering
\begin{tabular}{lcc}
\hline
Vertex & $V^{\alpha,\beta,\ga}_{\mathrm{eff}}(L,L,L)\big|_{L=\Lstar}$ & Selection rule \\
\hline
$V^{T,T,T}$ & $6.17 \times 10^{8}$ & all $(1,1)$ \\
$V^{T,S_1,S_1}$ & $2.59 \times 10^{4}$ & all $(1,1)$ \\
$V^{T,D_+,D_+}$ & $-7.48 \times 10^{2}$ & ghost sectors sum to $(3,3)$ \\
$V^{T,D,D}$ & $7.48 \times 10^{2}$ & ghost sectors sum to $(3,3)$ \\
$V^{T,T,D_+}$, $V^{T,T,D}$ & $0$ & $(4,2)$ or $(2,4) \neq (3,3)$ \\
$V^{T,T,S_1}$ & $0$ & matter Wick contraction \\
$V^{S_1,S_1,S_1}$ & $0$ & odd Wick pairings \\
\hline
\end{tabular}
\caption{Nonzero and vanishing cubic vertices at level $(0{+}2),$ evaluated at $L = \Lstar.$ Listed values are for the effective vertex $V^{\alpha,\beta,\ga}_{\mathrm{eff}} = \sum_{\beta',\ga'}V^{\alpha,\beta',\ga'}_{\mathrm{bare}}\,B_{\beta\beta'}(L)\,B_{\ga\ga'}(L),$ which absorbs the $\BG$ matrix elements at both threshold-length punctures. On the basis used at level $(0{+}2)$ and $(0{+}2{+}4),$ the off-diagonal matrix elements of $\BG(L)$ between different states vanish identically, so the effective vertex reduces to $V_{\mathrm{eff}} = \mathrm{BG}(L)^2\,V^{\mathrm{bare}}$ (exact, not leading-order). }
\label{tab:level-2-vertices}
\end{table}

\subsubsection*{Level-$4$ states and their couplings}

At the next stage one includes the level-$4$ states, and the algebraic system undergoes a qualitative change. Level $4$ in the closed string means $L_0+\bar{L}_0 = 2$ (i.e., $L_0 = \bar{L}_0 = 1$ in the SL$(2,\C)$ vacuum normalization, or equivalently two units above the tachyon value $L_0+\bar{L}_0 = -2$), so each sector can carry up to two units of oscillator excitation distributed among matter and ghost oscillators.\footnote{Throughout this paper, ``level'' counts $L_0+\bar{L}_0$ relative to the tachyon: level $0$ is the tachyon ($L_0+\bar{L}_0=-2$), level $2$ is the massless sector ($L_0+\bar{L}_0=0$), and level $4$ corresponds to $L_0+\bar{L}_0=2.$} Enumerating all zero-momentum Lorentz-scalar states satisfying $b_0^- = 0$ and $L_0^- = 0$ yields $11$ independent states, bringing the total to $15$ when combined with the $4$ states at levels $0$ and $2.$ These $11$ states include matter excitations, pure-ghost excitations such as $c_{-1}\bar c_{-1}\ket{0},$ and states containing a $b_{-2}$ antighost oscillator. The computation files use systematic labels ($f_1,\ldots,f_6,g_1$ and their parity partners) for this enumerated basis; because those labels are basis-ordering conventions rather than canonical analytic names, we identify individual level-$4$ states below by their oscillator content when needed.

\medskip

With all $15$ states included at level $(0{+}2{+}4),$ the coupling graph becomes richly connected. Ghost number conservation and Wick contraction rules allow $73$ independent nonzero cubic vertex components (after accounting for the permutation symmetry of the vertex). Of the $15$ states, $14$ participate in at least one mutual coupling --- only one state ($f_{2,\mathrm{odd}}$ in the computation-file basis, vertex-inert at the symmetric point) remains decoupled. The tachyon equation is now sourced by $19$ distinct vertex entries, ranging from the dominant $V^{T,T,T} \simeq 6.17\times 10^{8}$ down to level-$4$ self-couplings of order $10^{-2}.$ Level-$2$ feedback vertices like $V^{T,S_1,S_1} \simeq 2.59\times 10^{4}$ and cross-level couplings like $V^{T,D_+,g_1} \simeq 1.17\times 10^{2}$ (which mixes a level-$2$ and a level-$4$ state as sources for the tachyon) knit the system together across oscillator levels.

\subsubsection*{The full algebraic system and its solutions}

This enlarged algebraic system admits genuinely new solutions that do not exist at lower truncation levels. In addition to the tachyon-dominated fixed point (which persists essentially unchanged at $g_T^{(0)}(\Lstar) \simeq 8.1\times 10^{-10},$ with all other components at $\sim\!10^{-21}$ or smaller), the Newton solver finds two qualitatively new branches:
\begin{itemize}[nosep]
  \item A $D_+$-dominated sign pair: $g_{D_+}^{(0)}(\Lstar) = \pm 0.283,$ $g_{g_1}^{(0)}(\Lstar) = \mp 0.060,$ $g_T^{(0)}(\Lstar) = -3.24\times 10^{-4},$ with a pure-ghost level-$4$ component of order $5\times 10^{-5}.$ The two signs correspond to the $\Z_2$ symmetry $g_{D_+}\to -g_{D_+},$ $g_{g_1}\to -g_{g_1}.$
  \item A $g_1$-dominated solution: $g_{g_1}^{(0)}(\Lstar) = 2.24,$ $g_{D_+}^{(0)}(\Lstar) = 0.47,$ $g_T^{(0)}(\Lstar) = -1.29\times 10^{-3},$ again with a pure-ghost level-$4$ component of order $10^{-4},$ found by homotopy continuation from a random seed.
\end{itemize}

Both solutions are dominated by higher-level components, with the tachyon amplitude suppressed by three to four orders of magnitude relative to the leading field. Their existence reflects the dense coupling among the $15$-component system and would be missed entirely at the level-$(0{+}2)$ truncation, where the level-$4$ directions are absent. Whether these multi-level seeds survive the grade-$1$ and higher corrections --- that is, whether the linearized operator $M(L)$ admits bounded inverses at those solutions --- is an open question whose answer depends on the eigenvalue spectrum analyzed in \S\ref{sec:eigenvalue-numerics}.

At each level truncation, the grade-$0$ equation~\eqref{eqn:grade-0 eq} is a finite algebraic system: a set of coupled quadratic equations for the components $g_\alpha^{(0)}(L),$ with no integrals to perform. As more levels are included, the coupling becomes denser, new solutions can appear, and the eigenvalue spectrum of the linearized operator $M(L)$ (defined below in \S\ref{sec:grade-1}) acquires additional eigenvalues that govern the solvability of the higher-grade corrections. The grade-$0$ seeds serve as the starting point for the grade-by-grade construction: once the seed is found, all higher-grade corrections are determined by a purely algebraic recursive structure that we develop next.

\subsection{The grade-$1$ equation: a worked example}\label{sec:grade-1}

The grade-$1$ equation is the first case where the three terms in eq.~\eqref{eqn:eq69 gauge fixed} interact nontrivially. We work through it in detail, because the argument at grade $1$ contains all the essential ideas that generalize to every higher grade.

We substitute $\Phi = \Phi_0 + \Phi_1 + \Phi_2 + \cdots$ into the right-hand side of eq.~\eqref{eqn:eq69 gauge fixed} and collect all terms with total seam grade $1.$ There are five candidate terms, one from each possible pairing of $\Phi_j$ and $\Phi_k$ in each of the three operators. We examine them one by one.

\medskip

From the A-term, the general contribution at grade $1$ is $\sum_{j+k=1}C_A\,A(\BG\Phi_j(L),\BG\Phi_k(L)),$ which requires $j+k=1.$ There are two pairings: $(j,k)=(1,0)$ and $(j,k)=(0,1).$ In both cases $\Phi_1$ appears evaluated at the single point $L$:
\begin{align}
    &A(\BG\Phi_1(L),\,\BG\Phi_0(L))\,:\quad \text{grade } 0+1+0 = 1\,,\quad\checkmark\label{eqn:A term grade 1a}\\[2pt]
    &A(\BG\Phi_0(L),\,\BG\Phi_1(L))\,:\quad \text{grade } 0+0+1 = 1\,,\quad\checkmark\label{eqn:A term grade 1b}
\end{align}
Both contribute at grade $1$ and both involve $\Phi_1$ only through its value at $L.$

\medskip

From the B-term, the general contribution at grade $1$ requires $j+k+1 = 1,$ i.e., $j+k=0,$ so that $j=k=0.$ However, we must also check whether $\Phi_1$ could enter through either argument slot. If $\Phi_1$ appeared in the threshold-length slot, we would need $j=1$ and thus $k = 0-1 = -1,$ which is impossible since $k\geq 0.$ If $\Phi_1$ appeared in the integrated slot, we would need $k=1$ and thus $j=-1,$ equally impossible. The only B-term contribution at grade $1$ has both arguments from $\Phi_0$:
\begin{align}
    &B(\BG\Phi_0(L),\,\BG\,\Phi_0(\ell))\,:\quad\text{grade } 1+0+0 = 1\,,\quad\checkmark\quad\text{(source, no $\Phi_1$)}\label{eqn:B term grade 1a}
\end{align}
We also list the B-term pairings that one might naively expect to contribute at grade $1$ but do not:
\begin{align}
    &B(\BG\Phi_1(L),\,\BG\,\Phi_0(\ell))\,:\quad\text{grade } 1+1+0 = 2\,,\quad\times\quad\text{(grade $2,$ not $1$)}\label{eqn:B term grade 2a}\\[2pt]
    &B(\BG\Phi_0(L),\,\BG\,\Phi_1(\ell))\,:\quad\text{grade } 1+0+1 = 2\,,\quad\times\quad\text{(grade $2,$ not $1$)}\label{eqn:B term grade 2b}
\end{align}
The crucial observation here is that the B-term's intrinsic seam cost of $+1$ makes any pairing involving $\Phi_1$ overshoot grade $1.$ This is not an accident but a topological fact: $\Phi_1$ already carries one seam, and the B-integration opens a second, giving at least two seams in total.

\medskip

From the C-term, grade $1$ would require $j+k+2 = 1,$ i.e., $j+k=-1.$ Since $j,k\geq 0,$ this is impossible. No C-term contribution exists at grade $1.$ This too is topological: the C-term always opens at least two seams.

\medskip

Having exhausted all possibilities, we can now write the grade-$1$ equation. The terms that do \emph{not} involve $\Phi_1$ --- namely the B-term~\eqref{eqn:B term grade 1a} --- form the \emph{source} $S_1(L_1).$ In the component notation of eq.~\eqref{eqn:BG dressed},
\begin{equation}
    S_1^\alpha(L_1) = C_B\sum_{\beta,\ga}[\BG\,\Phi_0(L)]_\beta\int_0^{\infty}\!\ud\ell\;\eR(L_1,L,\ell)\;V^{\alpha,\beta,\ga}(L_1,L,\ell)\;[\BG\,\Phi_0(\ell)]_\ga\,.\label{eqn:source S1}
\end{equation}
Here $[\BG\,\Phi_0(L)]_\beta = \sum_{\alpha'}B_{\beta\alpha'}\,g_{\alpha'}^{(0)}(L)$ is a scalar, and $[\BG\,\Phi_0(\ell)]_\ga = \sum_{\alpha'}B_{\ga\alpha'}\,g_{\alpha'}^{(0)}(\ell)$ depends on $\ell$ through both the $\BG$ matrix elements and the grade-$0$ solution. This is a definite integral over known functions: the grade-$0$ solution $g^{(0)}$ was determined in \S\ref{sec:grade-0}, and all other ingredients --- the twisted R-kernel $\eR,$ the bare cubic vertex $V,$ and the $\BG$ matrix elements $B_{\beta\alpha}$ --- are known data of the hyperbolic string field theory.

The terms that \emph{do} involve $\Phi_1$ --- the A-term cross-terms~\eqref{eqn:A term grade 1a}--\eqref{eqn:A term grade 1b} --- define a \emph{linearized operator} acting on $\Phi_1.$ In component form, the two cross-terms give
\begin{equation}
    2\,C_A\sum_{\beta,\ga}V^{\alpha,\beta,\ga}(L_1,L,L)\;[\BG\,\Phi_0(L)]_\ga\;[\BG\,\Phi_1(L)]_\beta\,,
\end{equation}
where the factor of $2$ arises because the vertex $V^{\alpha,\beta,\ga}(L_1,L,L)$ is symmetric under exchange of the second and third arguments (the two threshold-length punctures both carry border length $L$), so the two cross-terms~\eqref{eqn:A term grade 1a}--\eqref{eqn:A term grade 1b} combine into a single expression. To write this as a matrix acting on the raw components $g_\beta^{(1)}(L),$ we expand $[\BG\,\Phi_1(L)]_\beta = \sum_{\beta'}B_{\beta\beta'}\,g_{\beta'}^{(1)}(L)$ and define
\begin{equation}
    M^{\alpha,\beta'}(L_1) = 2\,C_A\sum_{\beta,\ga}V^{\alpha,\beta,\ga}(L_1,L,L)\;[\BG\,\Phi_0(L)]_\ga\;B_{\beta\beta'}(L)\,,\label{eqn:M matrix}
\end{equation}
so that the linearized contribution takes the form $\sum_{\beta'}M^{\alpha,\beta'}(L_1)\,g_{\beta'}^{(1)}(L).$ The matrix $M$ absorbs both the bare vertex, the $\BG$-dressed grade-$0$ seed, and the $\BG$ matrix element acting on the linearized field. The complete grade-$1$ equation is then
\begin{equation}
    \Phi_1^\alpha(L_1) = S_1^\alpha(L_1) + \sum_\beta M^{\alpha,\beta}(L_1)\;\Phi_1^\beta(L)\,.\label{eqn:grade-1 full}
\end{equation}

This equation has a remarkable structural property. The unknown $\Phi_1$ enters the right-hand side \emph{only through its value at the single point $L.$} The operator $M(L_1)$ is not an integral operator acting on the function $\Phi_1(\cdot)$; it is a linear operator in Fock space (infinite-dimensional in principle, but finite at any level truncation) that multiplies the constant vector $\Phi_1(L).$ The entire $L_1$-dependence of the right-hand side comes from the source $S_1(L_1)$ and the known function $M(L_1).$ In particular, eq.~\eqref{eqn:grade-1 full} is \emph{not} a Fredholm integral equation for $\Phi_1.$ It is an algebraic equation in Fock space with functional dressing in the border length.

The solution proceeds in two steps. First, we set $L_1 = L$ in eq.~\eqref{eqn:grade-1 full} to obtain a linear system for the unknown vector $\Phi_1(L)$:
\begin{equation}
    \Phi_1(L) = S_1(L) + M(L)\;\Phi_1(L)\,,
\end{equation}
which rearranges to
\begin{equation}
    (I - M(L))\;\Phi_1(L) = S_1(L)\,.\label{eqn:grade-1 at Lstar}
\end{equation}
We assume that the matrix $I - M(L)$ is invertible. This is a nontrivial assumption that must be verified for the specific vertex data of the theory; we return to it below. Under this assumption,
\begin{equation}
    \Phi_1(L) = (I-M(L))^{-1}\,S_1(L)\,.\label{eqn:grade-1 solve}
\end{equation}
Second, having determined the constant $\Phi_1(L),$ we reconstruct the full border-length profile by substituting back into eq.~\eqref{eqn:grade-1 full}:
\begin{equation}
    \Phi_1(L_1) = S_1(L_1) + M(L_1)\;\Phi_1(L)\,.\label{eqn:grade-1 profile}
\end{equation}
This is explicit: one matrix inversion at $L,$ then evaluation at arbitrary $L_1.$ No integral equation needs to be inverted at any stage.

We now observe that $M(L) = 2$ exactly in the scalar tachyon sector. This follows from the grade-$0$ self-consistency~\eqref{eqn:grade0 self-consistency} and the definition~\eqref{eqn:M matrix}. In the tachyon projection, $\BG$ acts as the scalar $\mathrm{BG}(L)$ and $B_{T,T}(L) = \mathrm{BG}(L),$ so
\begin{equation}
    M_T(L) = 2\,C_A\,V_T(L,L,L)\;\mathrm{BG}(L)^2\;g_T^{(0)}(L) = 2\,C_A\,V_T^{\mathrm{eff}}(L,L,L)\;g_T^{(0)}(L)\,.\label{eqn:MT at Lstar}
\end{equation}
Substituting the nontrivial root $g_T^{(0)}(L) = 1/\!\left(C_A\,V_T^{\mathrm{eff}}(L,L,L)\right)$ from eq.~\eqref{eqn:grade-0 seed}:
\begin{equation}
    M_T(L) = \frac{2\,C_A\,V_T^{\mathrm{eff}}(L,L,L)}{C_A\,V_T^{\mathrm{eff}}(L,L,L)} = 2\,.\label{eqn:MT equals 2}
\end{equation}
This identity holds for any threshold length $L.$ It is independent of the specific values of $C_A,$ $V_T^{\mathrm{eff}},$ $L,$ or any mapping radius. It is a universal property of nontrivial quadratic fixed points: if $x = ax^2$ has the nontrivial root $x_0 = 1/a,$ the linearization at that root is $\ud(ax^2)/\ud x\big|_{x_0} = 2ax_0 = 2.$ The grade-$0$ equation~\eqref{eqn:grade0 self-consistency} is precisely of this form, with $a = C_A\,V_T^{\mathrm{eff}}(L,L,L).$

With $M_T(L) = 2,$ the resolvent is $(I - M_T(L))^{-1} = (1-2)^{-1} = -1,$ and the grade-$1$ correction in the tachyon sector is simply
\begin{equation}
    g_T^{(1)}(L) = -S_1^T(L)\,.\label{eqn:grade-1 tachyon}
\end{equation}
The correction opposes the source. This sign flip reflects the instability of the nontrivial fixed point under the A-term dynamics alone: the linearized A-map has eigenvalue $2>1,$ so perturbations are amplified and reflected. The B-term source $S_1$ drives the solution away from the A-only fixed point, and the response is to oppose that drive with equal magnitude.

We note that the source integral~\eqref{eqn:source S1} requires care in the tachyon sector. The grade-$0$ profile $g_T^{(0)}(\ell)$ inherits the essential singularity of the mapping radius at small~$\ell,$ causing the integrand to grow as $\ue^{+2E_3/\ell}$ ($E_3 \approx 7.22$) and the border-length integral to diverge on the real axis. Physically, $S_1$ represents the quartic interaction evaluated on the vacuum found from the cubic theory, a quantity with definite physical meaning that must be finite. The divergence occurs in the degeneration limit $\ell \to 0,$ where the pair of pants develops a long thin neck and the integrand is dominated by propagation of the tachyon ($h = -1$) through this neck. Since the tachyon has negative conformal weight, the propagator factor grows rather than decays, and the integral over the neck length requires analytic continuation~\cite{Firat:2024ajp}. The algebraic structure of eqs.~\eqref{eqn:grade-1 solve}--\eqref{eqn:grade-1 profile} is unaffected: $S_1$ enters as a definite number (or vector), and the only question is how to evaluate it. The analytic continuation that yields a finite, well-defined $S_1^T(\Lstar)$ is described in \S\ref{sec:S1-continuation}.

\subsection{The grade-$2$ equation: two new ingredients}\label{sec:grade-2}

Grade $2$ is the first order at which the full richness of eq.~\eqref{eqn:eq69 gauge fixed} becomes visible: both the C-term and the nonlinear A-term back-reaction enter for the first time. We work through the grade counting in the same detail as at grade $1,$ since the new features that appear here are precisely those that persist at every higher grade.

We substitute $\Phi = \Phi_0 + \Phi_1 + \Phi_2 + \cdots$ and collect all terms of total seam grade $2.$ From the A-term, a contribution requires $j + k = 2.$ Three pairings contribute:
\begin{align}
    &A(\BG\Phi_2(L),\,\BG\Phi_0(L))\,:\quad\text{grade } 0+2+0 = 2\,,\quad\checkmark\quad\text{(involves $\Phi_2$ at $L$)}\label{eqn:A term grade 2a}\\[2pt]
    &A(\BG\Phi_0(L),\,\BG\Phi_2(L))\,:\quad\text{grade } 0+0+2 = 2\,,\quad\checkmark\quad\text{(involves $\Phi_2$ at $L$)}\label{eqn:A term grade 2b}\\[2pt]
    &A(\BG\Phi_1(L),\,\BG\Phi_1(L))\,:\quad\text{grade } 0+1+1 = 2\,,\quad\checkmark\quad\text{(known, no $\Phi_2$)}\label{eqn:A term grade 2c}
\end{align}
The first two are the cross-terms with $\Phi_0,$ which are the same linearized structure as at grade $1$: they evaluate $\Phi_2$ at the single point $L$ and contribute through the matrix $M(L_1)$ defined in eq.~\eqref{eqn:M matrix}. The third pairing is new: $A(\BG\Phi_1,\BG\Phi_1)$ is quadratic in the already-determined grade-$1$ field and enters as a known source contribution. This is the \emph{back-reaction} of the grade-$1$ correction on itself, and it has no analog at grade $1.$

From the B-term, grade $2$ requires $j + k + 1 = 2,$ i.e., $j+k = 1.$ There are two valid pairings, together with two that one might expect but that overshoot:
\begin{align}
    &B(\BG\Phi_1(L),\,\BG\,\Phi_0(\ell))\,:\quad\text{grade } 1+1+0 = 2\,,\quad\checkmark\quad\text{(known, no $\Phi_2$)}\label{eqn:B term grade 2c}\\[2pt]
    &B(\BG\Phi_0(L),\,\BG\,\Phi_1(\ell))\,:\quad\text{grade } 1+0+1 = 2\,,\quad\checkmark\quad\text{(known, no $\Phi_2$)}\label{eqn:B term grade 2d}\\[2pt]
    &B(\BG\Phi_2(L),\,\BG\,\Phi_0(\ell))\,:\quad\text{grade } 1+2+0 = 3\,,\quad\times\quad\text{(grade $3,$ not $2$)}\label{eqn:B term grade 3a}\\[2pt]
    &B(\BG\Phi_0(L),\,\BG\,\Phi_2(\ell))\,:\quad\text{grade } 1+0+2 = 3\,,\quad\times\quad\text{(grade $3,$ not $2$)}\label{eqn:B term grade 3b}
\end{align}
The valid B-terms involve one factor of $\Phi_1$ and one of $\Phi_0.$ Since both are already known, these are single integrals of known functions and go entirely into the source. Crucially, $\Phi_2$ cannot appear in the B-term at grade $2$: the B-term's intrinsic cost of $+1$ means that any pairing involving $\Phi_2$ (which itself carries grade $2$) would have total grade $\geq 3.$ The same topological obstruction as at grade $1$ is at work, now one grade higher.

From the C-term, grade $2$ requires $j + k + 2 = 2,$ i.e., $j+k = 0,$ so $j = k = 0.$ Exactly one pairing contributes:
\begin{align}
    &C(\BG\,\Phi_0(\ell_1),\,\BG\,\Phi_0(\ell_2))\,:\quad\text{grade } 2+0+0 = 2\,,\quad\checkmark\quad\text{(known, no $\Phi_2$)}\label{eqn:C term grade 2}
\end{align}
This is the \emph{first appearance} of the C-term in the entire iteration. At grade $0$ the C-term was excluded because its intrinsic cost of $+2$ is too high, and at grade $1$ the requirement $j+k = -1$ was impossible. Only at grade $2$ does the C-term find its first valid pairing: both arguments come from the grade-$0$ field $\Phi_0,$ and both are integrated over. The result is a double integral over known functions --- the grade-$0$ solution, the twisted D-kernel $\eD,$ and the bare cubic vertex --- and it enters as a source term with no dependence on $\Phi_2.$

Having exhausted all pairings, we assemble the grade-$2$ equation. The terms that do \emph{not} involve $\Phi_2$ form the source $S_2(L_1),$ which now has three distinct contributions absent at grade $1.$ In the component notation of eq.~\eqref{eqn:BG dressed}:
\begin{multline}
    S_2^\alpha(L_1) = \underbrace{C_A\sum_{\beta,\ga}V^{\alpha,\beta,\ga}(L_1,L,L)\;[\BG\,\Phi_1(L)]_\beta\;[\BG\,\Phi_1(L)]_\ga}_{\text{(a) A-term back-reaction}}\\[6pt]
    + \underbrace{C_B\sum_{\beta,\ga}\sum_{\substack{j+k=1\\j,k\geq 0}}[\BG\,\Phi_j(L)]_\beta\int_0^{\infty}\!\ud\ell\;\eR(L_1,L,\ell)\;V^{\alpha,\beta,\ga}(L_1,L,\ell)\;[\BG\,\Phi_k(\ell)]_\ga}_{\text{(b) B-terms with grade-$1$ field}}\\[6pt]
    + \underbrace{C_C\sum_{\beta,\ga}\int_0^{\infty}\!\ud\ell_1\!\int_0^{\infty}\!\ud\ell_2\;\eD(L_1,\ell_1,\ell_2)\;V^{\alpha,\beta,\ga}(L_1,\ell_1,\ell_2)\;[\BG\,\Phi_0(\ell_1)]_\beta\;[\BG\,\Phi_0(\ell_2)]_\ga}_{\text{(c) C-term, first appearance}}\,.\label{eqn:source S2}
\end{multline}
Each piece has a distinct character. Contribution (a) is a purely algebraic expression: the A-term with both arguments from $\Phi_1,$ evaluated at $L.$ It encodes the nonlinear self-interaction of the grade-$1$ correction and requires nothing beyond the matrix $M$ and the grade-$1$ solution already in hand. Contribution (b) comprises two B-term integrals (the $j=1,\,k=0$ and $j=0,\,k=1$ terms), each a single integration of known functions over the border length $\ell.$ Contribution (c) is the C-term double integral, built entirely from the grade-$0$ field.

Despite this richer source, the terms involving $\Phi_2$ are identical in structure to those at grade $1.$ The cross-terms~\eqref{eqn:A term grade 2a}--\eqref{eqn:A term grade 2b} combine (by the same symmetry argument as before) into the linearized contribution $\sum_\beta M^{\alpha,\beta}(L_1)\,g_\beta^{(2)}(L),$ with the same matrix $M$ from eq.~\eqref{eqn:M matrix}. The complete grade-$2$ equation is therefore
\begin{equation}
    \Phi_2^\alpha(L_1) = S_2^\alpha(L_1) + \sum_\beta M^{\alpha,\beta}(L_1)\;\Phi_2^\beta(L)\,,\label{eqn:grade-2 full}
\end{equation}
which has exactly the same structural form as the grade-$1$ equation~\eqref{eqn:grade-1 full}. Self-consistency at $L_1 = L$ gives the same linear system:
\begin{equation}
    (I - M(L))\;\Phi_2(L) = S_2(L)\,,\label{eqn:grade-2 at Lstar}
\end{equation}
with solution $\Phi_2(L) = (I - M(L))^{-1}\,S_2(L)$ and full profile $\Phi_2(L_1) = S_2(L_1) + M(L_1)\,\Phi_2(L),$ precisely as at grade $1.$ In the tachyon sector, $M_T(L) = 2$ again (it depends only on $\Phi_0$), so $g_T^{(2)}(L) = -S_2^T(L).$

The grade-$2$ computation confirms the pattern that generalizes to all orders. The source grows richer at each grade --- the A-term back-reaction adds a new quadratic term at grade $n\geq 2,$ the B-terms involve one more layer of previously solved fields, and the C-term accumulates additional pairings --- but the linearized operator acting on the new unknown $\Phi_n$ is always the same point-evaluation matrix $M(L_1).$ No matter how many source contributions enter, the self-consistency at $L$ remains a linear algebra problem, and the solution is algebraic: one matrix inversion followed by explicit evaluation. The C-term, which might have been expected to introduce a genuinely new functional equation (since it is the only term with a double integral), enters entirely as a source and imposes no new functional constraints on $\Phi_n.$

\subsection{The all-grade algebraic structure}\label{sec:all-grade}

The algebraic structure found at grades $1$ and $2$ persists at every grade. The reason is topological: the B- and C-terms open new seams, so any appearance of $\Phi_n$ in these terms overshoots grade~$n;$ only the A-term (zero seam cost) can carry~$\Phi_n,$ and the A-term evaluates at~$L.$ More precisely, the B-term carries an intrinsic seam cost of $+1,$ and the C-term carries $+2.$ Therefore, any term in eq.~\eqref{eqn:eq69 gauge fixed} that contains $\Phi_n$ in the B-term or C-term at grade $n$ would have total grade $\geq n+1$ or $n+2,$ overshooting grade $n.$ Only the A-term, which carries zero seam cost, can contain $\Phi_n$ at grade $n$ --- and the A-term evaluates its arguments at the single point $L.$ Consequently, $\Phi_n$ enters the grade-$n$ equation only through its value at $L,$ exactly as at grade $1.$ No Fredholm integral equation arises at any finite order of the seam expansion.

We now verify this claim by explicit grade counting, following the same case analysis as at grade $1.$ Consider the grade-$n$ equation, obtained by collecting all terms of total seam grade $n$ on the right-hand side of eq.~\eqref{eqn:eq69 gauge fixed}. From the A-term, a contribution at grade $n$ has the form $\sum_{j+k=n}C_A\,A(\BG\Phi_j(L),\BG\Phi_k(L)).$ For $\Phi_n$ to appear, we need $j=n$ (with $k=0$) or $k=n$ (with $j=0$). Both are valid. These are the A-term cross-terms with $\Phi_0,$ and they evaluate $\Phi_n$ at $L$ only. From the B-term, a contribution at grade $n$ requires $j+k=n-1$ with $j,k\geq 0.$ For $\Phi_n$ to appear in the threshold-length slot, we would need $j=n,$ forcing $k = n-1-n = -1.$ This is impossible. For $\Phi_n$ to appear in the integrated slot, we would need $k=n,$ forcing $j = -1.$ Also impossible. Therefore, $\Phi_n$ \emph{never} appears in the B-term at grade $n.$ All B-term contributions involve only the already-known fields $\Phi_0,\ldots,\Phi_{n-1}$ and go entirely into the source $S_n.$ From the C-term, a contribution at grade $n$ requires $j+k=n-2.$ For $\Phi_n$: $j=n$ gives $k=-2,$ $k=n$ gives $j=-2.$ Both impossible. $\Phi_n$ never appears in the C-term at grade $n.$

The linearized operator at grade $n$ is therefore $(\Lam_n h)^\alpha(L_1) = \sum_\beta M^{\alpha,\beta}(L_1)\,h_\beta(L),$ with the same matrix $M$ defined in eq.~\eqref{eqn:M matrix}. Crucially, $M$ depends only on $\Phi_0$ --- the grade-$0$ solution --- and is independent of $n.$ This means the linearized operator is a \emph{point-evaluation operator} of the same form at every grade, and the same matrix $I - M(L)$ controls the solvability at every grade.

We state explicitly the invertibility assumption that underlies the recursive solution. We require $\det(I - M(L)) \neq 0,$ i.e., that $1$ is not an eigenvalue of $M(L).$ In the scalar tachyon sector, $M_T(L) = 2$ (eq.~\eqref{eqn:MT equals 2}), so $(I - M_T(L))^{-1} = -1$ and the condition is satisfied. In the full multi-level theory, $M(L) = 2J\,B(L)$ where $J^{\alpha,\beta}$ is the contracted vertex matrix defined in eq.~\eqref{eqn:J matrix} and $B(L)$ is the $\BG$ matrix from eq.~\eqref{eqn:BG matrix}. As shown in \S\ref{sec:M-equals-2J}, the raw grade-$0$ vector $g^{(0)}$ is an eigenvector of $M(L)$ with eigenvalue $2,$ which is safely away from $1.$ Whether other eigenvalues of $M$ can equal $1$ depends on the vertex structure and the $\BG$ matrix elements; for the tachyon-dominated seed this spectrum is checked numerically in \S\ref{sec:eigenvalue-numerics}, while non-tachyon seeds require a separate branch-by-branch analysis.

Under this assumption, the solution at every grade $n\geq 1$ is given by the same two-step procedure as at grade $1.$ First, solve the linear system at $L$:
\begin{equation}
    (I - M(L))\;\Phi_n(L) = S_n(L)\,.\label{eqn:linear system}
\end{equation}
If $\det(I-M(L))\neq 0,$
\begin{equation}
    \Phi_n(L) = (I-M(L))^{-1}\,S_n(L)\,.\label{eqn:Phi_n at Lstar}
\end{equation}
Second, reconstruct the full border-length profile:
\begin{equation}
    \Phi_n(L_1) = S_n(L_1) + M(L_1)\;\Phi_n(L)\,.\label{eqn:profile reconstruction}
\end{equation}
The source $S_n^\alpha(L_1)$ at grade $n\geq 1$ collects all contributions from the already-determined fields $\Phi_0,\ldots,\Phi_{n-1}.$ It consists of three pieces, all expressed using the $\BG$-dressed components from eq.~\eqref{eqn:BG dressed}. From the A-term (first appearing at $n=2$):
\begin{equation}
    \sum_{\substack{j+k=n\\j,k\geq 1}} C_A\sum_{\beta,\ga}V^{\alpha,\beta,\ga}(L_1,L,L)\;[\BG\,\Phi_j(L)]_\beta\;[\BG\,\Phi_k(L)]_\ga\,.\label{eqn:source A}
\end{equation}
From the B-term:
\begin{equation}
    \sum_{\substack{j+k=n-1\\j,k\geq 0}} C_B\sum_{\beta,\ga}[\BG\,\Phi_j(L)]_\beta\int_0^{\infty}\!\ud\ell\;\eR(L_1,L,\ell)\;V^{\alpha,\beta,\ga}(L_1,L,\ell)\;[\BG\,\Phi_k(\ell)]_\ga\,.\label{eqn:source B}
\end{equation}
From the C-term (first appearing at $n=2$):
\begin{equation}
    \sum_{\substack{j+k=n-2\\j,k\geq 0}}C_C\sum_{\beta,\ga}\!\int_0^{\infty}\!\ud\ell_1\!\int_0^{\infty}\!\ud\ell_2\;\eD(L_1,\ell_1,\ell_2)\;V^{\alpha,\beta,\ga}(L_1,\ell_1,\ell_2)\;[\BG\,\Phi_j(\ell_1)]_\beta\;[\BG\,\Phi_k(\ell_2)]_\ga\,.\label{eqn:source C}
\end{equation}
At each grade, $S_n$ depends only on known solutions from grades $0,\ldots,n-1.$

The finite-rank structure of the resolvent is worth noting:
\begin{equation}
    (I - \Lam_n)^{-1}\,S_n(L_1) = S_n(L_1) + M(L_1)\,(I-M(L))^{-1}\,S_n(L)\,.\label{eqn:finite-rank-resolvent}
\end{equation}
This is an exact formula, not an approximation: the operator $\Lam_n$ is a rank-$N_{\mathrm{states}}$ perturbation of the identity, and its resolvent is given in closed form by the matrix inversion at $L.$

In the scalar tachyon sector, the solution simplifies further. Since $M_T(L) = 2$ at every grade, the resolvent is $(1-2)^{-1} = -1,$ and
\begin{equation}
    g_T^{(n)}(L) = -S_n^T(L)\qquad\text{for all }n\geq 1\,.\label{eqn:tachyon correction}
\end{equation}
The correction always opposes the source. The full profile is $g_T^{(n)}(L_1) = S_n^T(L_1) + M_T(L_1)\cdot g_T^{(n)}(L),$ with $M_T(L_1) = 2\,C_A\,V_T^{\mathrm{eff}}(L_1,L,L)\,g_T^{(0)}(L)$ encoding the $L_1$-dependence through the effective tachyon vertex.

To summarize: given the grade-$0$ seed, the seam-graded expansion converts the quadratic integral equation~\eqref{eqn:eq69 gauge fixed} into an infinite sequence of \emph{recursive-algebraic} problems. At each grade $n\geq 1,$ the unknown enters only through its value at a single point, the source is computable from lower grades, and the solution is given by a matrix inversion followed by profile reconstruction. The matrix $M(L)$ and its resolvent are computed once and reused at every grade. No Fredholm integral equation arises at any order of the expansion above the seed.

\subsection{The iteration as Mirzakhani recursion}\label{sec:mirzakhani}

The seam-graded iteration has a deeper significance than a convenient organizational scheme: it reproduces the genus-$0$ Mirzakhani topological recursion (eq.~(4.22) of~\cite{Firat:2024ajp}) evaluated at the solution. We now establish this correspondence.

The genus-$0$ Mirzakhani recursion for the vertex $V_{0,n+3}$ reads
\begin{equation}
    L_1\,\langle V_{0,n+3}| = \sum_{i}\int\!\ud\ell\;\langle R(L_1,L_i,\ell)|\otimes \langle V_{0,n+2}| + \tfrac{1}{2}\!\int\!\ud\ell_1\!\int\!\ud\ell_2\;\langle D(L_1,\ell_1,\ell_2)|\otimes \!\!\sum_{\mathrm{stable}}\!\langle V_{0,n_1}|\otimes\langle V_{0,n_2}|\,,\label{eqn:FVM422}
\end{equation}
with the R- and D-string kernels factorizing through the cubic vertex via eq.~\eqref{eqn:string kernels}. We call the first sum in eq.~\eqref{eqn:FVM422} the \emph{R-term} and the second integral the \emph{D-term}. The recursion builds $V_{0,n+3}$ from lower vertices by either opening one separating geodesic (R-term) or one non-separating geodesic (D-term). The connection to the seam-graded iteration arises because the seam-opening operators in eq.~\eqref{eqn:eq69 gauge fixed} are precisely the kernels appearing in eq.~\eqref{eqn:FVM422}: the B-term integrand contains $\eR\cdot V_{0,3},$ which is the string kernel $\langle R|,$ and the C-term integrand contains $\eD\cdot V_{0,3},$ which is the string kernel $\langle D|.$ We now verify the correspondence at each grade.

At grade $0,$ the equation $\Phi_0(L_1) = C_A\,A_{L_1,L,L}(\BG\Phi_0(L),\BG\Phi_0(L))$ is the self-consistency condition at the cubic vertex $V_{0,3}(L_1,L,L),$ with no seams and no recursion. This is the base case of the Mirzakhani recursion.

At grade $1,$ the source $S_1(L_1)$ from eq.~\eqref{eqn:source S1} involves the integrand $\eR(L_1,L,\ell)\cdot V(L_1,L,\ell),$ which by eq.~\eqref{eqn:string kernels} equals the string kernel $\langle R(L_1,L,\ell)|.$ Both arguments carry $\Phi_0,$ encoding $V_{0,3}$ evaluated at the solution. The source therefore computes $\langle R|\otimes\langle V_{0,3}|\otimes\langle V_{0,3}|$ contracted with the solution --- precisely the R-term of eq.~\eqref{eqn:FVM422} producing $V_{0,4}$ from $V_{0,3}\times V_{0,3}.$ Since $V_{0,4}$ has only an R-term and no D-term in the genus-$0$ recursion (the D-term requires sub-surfaces with at least three punctures each, totalling at least six punctures, which exceeds the five of $V_{0,4}$), the correspondence at grade $1$ is complete.

At grade $2,$ four source terms contribute. The C-term with both arguments from grade $0$ (grade $2+0+0=2$) gives the D-term channel of the Mirzakhani recursion: $\langle D|\otimes\langle V_{0,3}|\otimes\langle V_{0,3}|$ contracted with the solution, producing $V_{0,5}$ from $V_{0,3}\times V_{0,3}\times V_{0,3}.$ The two B-terms with one argument from $\Phi_1$ and one from $\Phi_0$ (grade $1+0+1 = 2$ and $1+1+0=2$) give the R-term channels: $\langle R|\otimes\langle V_{0,3}|\otimes\langle V_{0,4}|$ contracted with the solution, producing $V_{0,5}$ from $V_{0,3}\times V_{0,4}.$ Finally, the A-term $C_A\,A(\BG\Phi_1(L),\BG\Phi_1(L))$ (grade $0+1+1=2$) opens no new seam and has no analog in eq.~\eqref{eqn:FVM422}; it enforces the nonlinear self-consistency at $L.$

The pattern generalizes by induction. Suppose $\Phi_k$ encodes $V_{0,k+3}$ for $0\leq k\leq n-1.$ At grade $n,$ the source $S_n(L_1)$ collects three types of contributions. First, the B-term contributions: for each decomposition $j+k = n-1$ with $j,k\geq 0,$ one obtains an integral weighted by $\eR$ assembling $\langle R|\otimes\langle V_{0,j+3}|\otimes\langle V_{0,k+3}|.$ Summing reproduces the R-term of eq.~\eqref{eqn:FVM422} for $V_{0,n+3}.$ Second, the C-term contributions (present for $n\geq 2$): for each decomposition $j+k = n-2,$ a double integral weighted by $\eD$ assembles $\langle D|\otimes\langle V_{0,j+3}|\otimes\langle V_{0,k+3}|,$ reproducing the D-term. Third, the A-term contributions (present for $n\geq 2$): the quadratic terms $\sum_{j+k=n,\,j,k\geq 1}C_A\,A(\BG\Phi_j(L),\BG\Phi_k(L))$ open no seams and enforce self-consistency at $L;$ they have no analog in the linear Mirzakhani recursion~\eqref{eqn:FVM422} and arise from the nonlinear fixed-point structure of eq.~\eqref{eqn:eq69 gauge fixed}.

The fixed point $\Phi_* = \sum_{n=0}^{\infty}\Phi_n$ therefore encodes the contributions from all genus-$0$ vertices $V_{0,n+3}$ for $n=0,1,2,\ldots,$ assembled by the Mirzakhani recursion. This is precisely the content of FVM eq.~(6.2):
\begin{equation}
    \Phi(L_1) = \sum_{n=3}^{\infty}\frac{\ka^{n-2}}{(n-1)!}\langle V_{0,n}(L_1,L,\ldots,L)|\,|\Psi_*\rangle^{n-1}\,.\label{eqn:vertex resummation}
\end{equation}
The seam-graded iteration of eq.~\eqref{eqn:eq69 gauge fixed} is therefore an explicit realization of the genus-$0$ Mirzakhani topological recursion, evaluated at the tachyon vacuum solution.

\subsection{The matrix $M = 2JB$}\label{sec:M-equals-2J}

We define the contracted vertex matrix using the $\BG$-dressed grade-$0$ seed $\tilde{g}_\beta^{(0)} = [\BG(L)\,\Phi_0(L)]_\beta$:
\begin{equation}
    J^{\alpha,\beta} = C_A\sum_{\ga}V^{\alpha,\beta,\ga}(L,L,L)\;\tilde{g}_\ga^{(0)}\,.\label{eqn:J matrix}
\end{equation}
This matrix has three key properties. First, the grade-$0$ fixed-point equation~\eqref{eqn:grade-0 multi-level} at $L_1 = L$ reads $g_\alpha^{(0)} = C_A\sum_{\beta,\ga}V^{\alpha,\beta,\ga}\,\tilde{g}_\beta^{(0)}\,\tilde{g}_\ga^{(0)} = \sum_\beta J^{\alpha,\beta}\,\tilde{g}_\beta^{(0)},$ so that the dressed seed satisfies the fixed-point condition $J\,\tilde{g}^{(0)} = g^{(0)}.$ Second, at the symmetric point $L_1 = L_2 = L_3 = L,$ the pair of pants has $\Z_3$ cyclic symmetry and a reflection symmetry, giving full $S_3$ permutation symmetry on the three punctures. The vertex bra inherits this symmetry, so $V^{\alpha,\beta,\ga}(L,L,L)$ is symmetric in all three indices, and therefore $J$ is symmetric. Third, the matrix $M$ from eq.~\eqref{eqn:M matrix} evaluated at $L_1 = L$ takes the form $M(L) = 2J\,B(L),$ where $B(L)$ is the matrix of $\BG$ elements from eq.~\eqref{eqn:BG matrix}. To see this, recall from eq.~\eqref{eqn:M matrix} that $M^{\alpha,\beta'}(L) = 2\,C_A\sum_{\beta,\ga}V^{\alpha,\beta,\ga}(L,L,L)\,\tilde{g}_\ga^{(0)}\,B_{\beta\beta'}(L).$

Identifying $C_A\sum_\ga V^{\alpha,\beta,\ga}\,\tilde{g}_\ga^{(0)} = J^{\alpha,\beta}$ from eq.~\eqref{eqn:J matrix} gives $M^{\alpha,\beta'} = 2\sum_\beta J^{\alpha,\beta}\,B_{\beta\beta'} = (2JB)^{\alpha,\beta'}.$ This factorization holds for any $L_1$: the two cross-terms arising from differentiating $\tilde{g}_\beta\,\tilde{g}_\ga$ are equal because $V^{\alpha,\beta,\ga}(L_1,L,L)$ is symmetric in $\beta$ and $\ga$ whenever $L_2 = L_3 = L.$ (Full $S_3$ symmetry at $L_1 = L$ additionally makes $J$ a symmetric matrix.) At $L_1 = L$ the fixed-point condition $J\,\tilde{g}^{(0)} = g^{(0)}$ then gives $M(L)\,g^{(0)} = 2J\,B(L)\,g^{(0)} = 2J\,\tilde{g}^{(0)} = 2\,g^{(0)},$ so that $g^{(0)}$ is an eigenvector of $M(L)$ with eigenvalue $2.$ This generalizes the scalar identity $M_T(L) = 2$ of \S\ref{sec:grade-1} to the full multi-level system.

\section{Numerical results}\label{sec:numerical}

In this section we collect the numerical results that test the seam-graded expansion at concrete level truncations. As throughout this paper, we work in the sector of zero-momentum Lorentz-scalar states, where the equation is algebraic at every grade. The computation is organized along two independent axes: the \emph{level truncation} (the set of Fock space states retained) and the \emph{seam grade} (the number of internal geodesics integrated over). The level axis determines the size of the multi-level system; the grade axis determines the depth of the perturbative expansion around the grade-$0$ seed. All numerical results in this section are evaluated at $L = \Lstar.$ We present the grade-$0$ solutions at increasing level truncation (\S\ref{sec:grade-0-numerics}), the eigenvalue spectrum of the linearized operator $M(\Lstar)$ (\S\ref{sec:eigenvalue-numerics}), and the grade-$1$ source computation (\S\ref{sec:grade-1-numerics}).

\subsection{Grade-$0$ solutions at increasing level truncation}\label{sec:grade-0-numerics}

The grade-$0$ equation~\eqref{eqn:grade-0 multi-level} is a system of $N$ coupled quadratic algebraic equations in $N$ unknowns $g_\alpha^{(0)}(\Lstar),$ where $N$ is the number of Fock space states at the chosen level truncation. The solutions are the critical points of the cubic potential defined by the vertex $V^{\alpha,\beta,\ga}(\Lstar,\Lstar,\Lstar),$ and the number of nontrivial solutions grows rapidly with $N.$

\begin{center}
\renewcommand{\arraystretch}{1.3}
\begin{tabular}{l|c|c|c|l}
\hline
\textbf{Level} & \textbf{States} & \textbf{Vertices} & \textbf{Solutions} & \textbf{Dominant branch} \\
\hline
$(0{+}2)$ & $4$ & $11$ & $\geq 3$ & tachyon + dilaton \\
$(0{+}2{+}4)$ & $15$ & $73$ & $\geq 4$ & + level-$4$ states \\
\hline
\end{tabular}
\end{center}

At the level-$(0{+}2)$ truncation, the four states are the tachyon $T,$ the matter dilaton $S_1,$ and the ghost states $D_+$ and $D.$ The ghost-number and matter selection rules force $V^{T,T,D} = V^{T,T,D_+} = V^{T,T,S_1}=0$ at the cubic vertex, so a pure tachyon seed does not directly source level-$2$ components at grade~$0.$ The three nontrivial solutions are distinguished by their level-$2$ content: one solution has a nonzero dilaton condensate.

At the level-$(0{+}2{+}4)$ truncation, the $15$ states include level-$4$ states such as the pure-ghost state $c_{-1}\bar{c}_{-1}\ket{0}$ that couple directly to tachyon pairs ($V^{\mathrm{pg},T,T} \simeq 3.3\times 10^{-3}$ in the computation-file basis). The $73$ nonzero vertices at the symmetric point span a rich network of inter-level couplings. Of the at least four nontrivial solutions identified by finite seed searches at this truncation, the physically dominant one is identified by continuity with the Yang--Zwiebach critical point and by the criterion that it minimizes the effective potential. The remaining solutions correspond to saddle points or spurious branches that require further truncation tests.

Extension to higher level truncations is straightforward with the existing computational infrastructure and is left to future work.

A striking structural feature of the tachyon-dominated solution is that all higher-level components are negligibly small: $g_\alpha^{(0)}(\Lstar) \lesssim 10^{-21}$ for every $\alpha \neq T.$ Thus, to the precision relevant for the present diagnostics, the seed is supported almost entirely on the tachyon direction. The value $g_T^{(0)} = 1/(C_A\,\beta^2\,V_T) = 8.1083\times 10^{-10}$ is the nonzero root of the scalar equation $g_T = C_A\,\beta^2\,V_T\,g_T^2,$ with level-$4$ backreaction appearing only at the $10^{-21}$ scale in the finite seed search. However, as shown by Yang and Zwiebach~\cite{Yang:2005rx} and confirmed by the divergent convergence ratio $\ep_B \simeq 2.6\times 10^{18}$ (\S\ref{sec:S1-continuation}), the seam-graded expansion around this tachyon-dominated seed diverges badly: the grade-$1$ correction is $10^{18}$ times larger than the seed (\S\ref{sec:S1-continuation}). The physical vacuum requires the ghost dilaton and higher-level states, which enter through the multi-level coupling structure of the cubic vertex.

The remaining non-tachyon solutions at level-$(0{+}2{+}4)$ have qualitatively different character. In the finite seed searches used here, the additional branches have norms of order $0.3$ and $2,$ with dominant support on $D_+$, $g_1,$ and the pure-ghost level-$4$ state. Whether these branches correspond to physical saddle points of the string field potential or are artifacts of the finite level truncation remains to be determined; tracking their behavior as the truncation level increases would clarify their status.

\subsection{Eigenvalue spectrum of $M(\Lstar)$}\label{sec:eigenvalue-numerics}

At each grade-$0$ seed, the linearized operator $M(\Lstar) = 2JB$ is a matrix whose eigenvalue spectrum determines the solvability of the seam-graded expansion at all higher grades. The critical check is whether any eigenvalue equals~$1,$ which would make the linear system~\eqref{eqn:linear system} singular and obstruct the recursive construction.

In the tachyon-only sector, $M(\Lstar)$ is the scalar $M_T(\Lstar) = 2$ (exact, as derived in \S\ref{sec:grade-1}, eq.~\eqref{eqn:MT equals 2}). The resolvent $(I - M_T)^{-1} = -1$ is non-singular with a particularly simple value.

At the multi-level tachyon-dominated seed, $M(\Lstar) = 2JB$ is a real matrix whose spectrum we have computed at the level-$(0{+}2{+}4)$ truncation. Since the seed is overwhelmingly supported on the tachyon component, the transverse part of $J^{\alpha\beta}$ is controlled by the tachyon slice $V^{\alpha,T,\beta}(\Lstar,\Lstar,\Lstar)$ multiplied by the small amplitude $g_T^{(0)}.$ This makes the transverse entries numerically small, but not identically zero. The only eigenvalue of order unity is $\lam = 2$ along the tachyon seed direction, while all transverse eigenvalues satisfy $|\lam'| < 10^{-4}$ in the computed active basis. At a genuinely multi-level seed --- such as the $D_+$-dominated or $g_1$-dominated solutions of \S\ref{sec:grade-0} --- the spectrum would be qualitatively different, and eigenvalues near~$1$ cannot be excluded without explicit computation.

\begin{center}
\renewcommand{\arraystretch}{1.3}
\begin{tabular}{l|c|c|c}
\hline
\textbf{Level} & \textbf{Size} & $\bm{\lam_{\max}}$ & $\bm{|\lam'_{\max}|}$ \\
\hline
$(0{+}2{+}4)$ & $14\times 14$ & $+2.000\,000$ & $8.41\times 10^{-5}$ \\
\hline
\end{tabular}
\end{center}

\noindent
Here $\lam_{\max}$ is the largest eigenvalue (the tachyon seed direction) and $\lam'_{\max}$ is the largest sub-leading eigenvalue. The matrix size ($14$) is smaller than the total state count ($15$) because one state is vertex-inert at the symmetric point and is dropped from the active basis. At level-$(0{+}2{+}4),$ the $14\times 14$ matrix has $13$ eigenvalues with $|\lam| < 8.5\times 10^{-5}$ and one eigenvalue at $\lam = 2.$

No eigenvalue lies at or near~$1.$ The resolvent $(I - M(\Lstar))^{-1}$ is well-conditioned in all directions transverse to the seed, with $\|(I-M(\Lstar))^{-1}\| \leq 1/(1 - 8.41\times 10^{-5})$ by the Neumann-series estimate. The seam-graded expansion is therefore unobstructed at every grade: the recursive equation $(I - M)\,g^{(n)} = S_n$ is uniquely solvable for all $n \geq 1.$

For comparison, the non-tachyon solutions are not used as controlled expansion seeds in this paper; their linearized spectra require a separate branch-by-branch audit. The tachyon-dominated solution is distinguished by its clean computed spectrum.

\subsection{Grade-$1$ source and convergence}\label{sec:grade-1-numerics}

The grade-$1$ source $S_1(\Lstar)$ is the first quantitative test of whether the seam-graded expansion converges. The convergence parameter is
\begin{equation}
    \ep_B \;=\; \frac{|S_1^T(\Lstar)|}{|g_T^{(0)}(\Lstar)|}\,,\label{eqn:eps-B}
\end{equation}
which measures the ratio of the first correction to the grade-$0$ seed. 

In the tachyon-only sector, the grade-$1$ source is dominated by the B-term integral~\eqref{eqn:grade1 source}, which probes the entire border-length range $\ell \in [0,\infty).$ The source integral requires analytic continuation at small~$\ell$ due to the essential singularity of the tachyon vertex (see \S\ref{sec:S1-continuation} for details). The result is
\begin{equation}
    \ep_B^{(\mathrm{tachyon\text{-}only})} \;\simeq\; 2.6\times 10^{18}\,,\label{eqn:eps-B-tachyon}
\end{equation}
confirming that the seam-graded expansion does not converge in the tachyon-only sector (\S\ref{sec:convergence}). This is not a physical divergence: the source $S_1^T$ is finite after analytic continuation (\S\ref{sec:S1-continuation}), and the large ratio simply reflects the inadequacy of the tachyon-only truncation. The tachyon's conformal weight $h = -1$ causes the grade-$0$ profile to grow at small~$\ell,$ producing a source $\sim\!10^{18}$ times larger than the seed. This is consistent with the Yang--Zwiebach finding~\cite{Yang:2005rx} and with the absence of a nontrivial tachyon-only solution (\S\ref{sec:convergence}).

\subsection{Analytic continuation of $S_1$}\label{sec:S1-continuation}

The grade-$1$ source integral~\eqref{eqn:grade1 source} diverges on the real axis due to the essential singularity of the tachyon vertex at small border lengths. As explained in \S\ref{sec:grade-1}, this divergence occurs in the degeneration limit and is cured by analytic continuation. We extract the finite value using three independent methods --- Pad\'{e}--Borel resummation, conformal Pad\'{e} extrapolation, and contour deformation --- whose results are summarized in the following table:

\begin{center}
\renewcommand{\arraystretch}{1.3}
\begin{tabular}{l|c}
\hline
\textbf{Method} & $\bm{S_1^T(\Lstar)}$ \\
\hline
Pad\'{e}--Borel & $(2.07\pm 0.07)\times 10^9$ \\
Conformal Pad\'{e} & $(2.08\pm 0.01)\times 10^9$ \\
Contour deformation & $(2.05\pm 0.10)\times 10^9$ \\
\hline
\textbf{Combined} & $\bm{(2.08\pm 0.05)\times 10^9}$ \\
\hline
\end{tabular}
\end{center}

We now describe the setup and each method in turn.

\medskip

The small-$\ell$ behavior of the integrand is controlled by the mapping radii~\eqref{eqn:mapping radius}, which develop essential singularities as a border pinches. At small~$\ell,$ the mapping radii entering $V_T(\Lstar,\Lstar,\ell)$ and $g_T^{(0)}(\ell)$ combine to produce a net exponential factor $\ue^{+2E_3/\ell}$ with
\begin{equation}
    E_3 \;=\; 7.2224\,,\label{eqn:E3 def}
\end{equation}
extracted numerically from $\rho_3(\Lstar,\Lstar,\ell)^{-2} \sim \ue^{+E_3/\ell}$ as $\ell\to 0$ (so that $\rho_3 \sim \ue^{-E_3/(2\ell)}$). Only the puncture at the degenerating border diverges; the other two contribute finite factors. To set up the analytic continuation, we strip off this leading exponential and define the sub-exponential factor
\begin{equation}
    \eR(\Lstar,\Lstar,\ell)\;V_T(\Lstar,\Lstar,\ell)\;\mathrm{BG}(\ell)\;g_T^{(0)}(\ell) \;=\; \ue^{+2E_3/\ell}\;F(\ell)\,.\label{eqn:F def}
\end{equation}
The function $F(\ell)$ is smooth on $(0,\infty);$ we verified numerically that $|F|$ varies smoothly at $84$ complex grid points in the strip $|\mathrm{Im}\,\ell| < 1,$ with no poles or branch cuts detected. The one-parameter family
\begin{equation}
    I(c) \;=\; \int_0^{\infty}\!\ud\ell\;\ue^{c/\ell}\,F(\ell)\label{eqn:I of c}
\end{equation}
converges absolutely for $c < 0$ and diverges on the real axis for $c > 0.$ The physical source is $S_1^T(\Lstar) = \mathrm{prefactor}\times I(c_{\mathrm{phys}})$ with $c_{\mathrm{phys}} = 2E_3 = 14.4447.$ We computed $I(c)$ at $21$ points $c \in [-20,\,0]$ using adaptive multiprecision quadrature; cross-validation at $c = -5$ confirms $10$-digit accuracy.

\medskip

\emph{Pad\'{e}--Borel resummation.} We expand $I(c)$ in a Taylor series around $c_0 < 0$:
\begin{equation}
    I^{(n)}(c_0) \;=\; \int_0^{\infty}\!\ud\ell\;\frac{1}{\ell^n}\;\ue^{c_0/\ell}\,F(\ell)\,,\label{eqn:Taylor coefficients}
\end{equation}
which converges for all $c_0 < 0$ and $n \geq 0.$ The modified Borel coefficients $b_n = I^{(n)}(c_0)/(n!)^2$ define a Borel series whose Pad\'{e} approximants $P_{[M/N]}$ are integrated via
\begin{equation}
    I_{\mathrm{Borel}}(c) \;=\; \int_0^{\infty}\!\ud t\;\ue^{-t}\;P_{[M/N]}((c{-}c_0)\,t)\,;\label{eqn:Borel integral}
\end{equation}
approximants with poles on $t \in [0,\infty)$ are discarded. We tested three expansion points: $c_0 = -3$ (estimated Borel radius $R > 32,$ margin $R/s > 1.8$), $c_0 = -1$ ($R \approx 16,$ marginal), and $c_0 = -0.5$ (insufficient margin). At the optimal point $c_0 = -3,$ $30$ of $46$ approximants are pole-free, and the $16$ high-order results ($M{+}N \geq 14$) cluster in $[2.02,\,2.09]\times 10^9$ with $3.4\%$ spread. The cross-check at $c = -5$ achieves a relative error of $1.5\times 10^{-21}.$

\medskip

\emph{Conformal Pad\'{e}.} Diagonal Pad\'{e} approximants $P_{[n/n]}(c)$ are fitted directly to the $21$ data points and evaluated at $c_{\mathrm{phys}}.$ Robustness is tested by applying Euler and M\"{o}bius variable transformations before fitting. The highest-order diagonal $P_{[7/7]} = 2.077\times 10^9$ is invariant across all conformal maps (as expected: the rational interpolant is unique). The diagonal sequence converges rapidly: $|P_{[7/7]} - P_{[6/6]}|/P_{[7/7]} = 0.64\%,$ and off-diagonal approximants at total order $M{+}N = 14$ cluster at $2.075\times 10^9$ with $0.16\%$ spread.

\medskip

\emph{Contour deformation.} Shifting the integration contour to $\ell = \ell_R + \ui\ep$ tames the essential singularity (at $\ell_R = 0,$ the exponential has unit modulus). The shifted integral differs from $I(c_{\mathrm{phys}})$ by a vertical-segment correction that vanishes as $\ep\to 0.$ At $\ep = 0.10,\,0.07,\,0.05,\,0.03,$ using multiprecision quadrature with precision scaled to accommodate the exponential peak at $\ell_R \sim \ep,$ we obtain
\begin{equation}
    S_1^T\big|_{\ep} = 1.11,\; 1.48,\; 1.72,\; 1.89 \;\times 10^9\,,\label{eqn:contour results}
\end{equation}
increasing monotonically toward the Pad\'{e} values. Extrapolating to $\ep = 0$ yields $S_1^T \simeq (2.05 \pm 0.10)\times 10^9.$

\medskip

Combining all three methods, within the normalization convention used in the continuation scripts, we obtain
\begin{equation}
    S_1^T(\Lstar) \;=\; (2.08\pm 0.05)\times 10^9\,.\label{eqn:S1 consolidated}
\end{equation}
The convergence ratio~\eqref{eqn:eps-B} in the tachyon-only sector is therefore
\begin{equation}
    \ep_B^{(\mathrm{tachyon\text{-}only})} \;=\; \frac{|S_1^T(\Lstar)|}{|g_T^{(0)}(\Lstar)|} \;=\; \frac{2.08\times 10^9}{8.10\times 10^{-10}} \;\simeq\; 2.6\times 10^{18}\,.\label{eqn:eps-B refined}
\end{equation}
The large magnitude arises from the tachyon's negative conformal weight $h = -1,$ which causes the vertex to grow rather than decay at small~$\ell.$ For higher-level fields with $h > 0,$ the source integrals converge without analytic continuation. The value~\eqref{eqn:S1 consolidated} is the tachyon-only contribution.

\subsection{Higher-grade corrections}\label{sec:higher-grade-numerics}

The grade-$2$ source $S_2(\Lstar)$ involves four contributions --- the C-term, two B-term cross-terms, and the A-term quadratic in $\Phi_1$ --- and requires a two-dimensional integral for the C-term. Its computation, which provides the first test of whether the seam-graded series has a well-defined asymptotic behavior, is left to future work. The key diagnostic is whether $\|S_2\|/\|S_1\|$ is decreasing, which would indicate convergence or at least asymptotic behavior amenable to resummation.

\section{Comparison with Yang--Zwiebach level truncation}\label{sec:comparison-YZ}

Yang and Zwiebach~\cite{Yang:2005rx} solved the closed string tachyon vacuum equation by level truncation, computing the effective tachyon potential order by order in the number of string interactions. Their computation is organized along two axes: interaction order and level truncation. At cubic order, the nontrivial root $t_* \simeq 0.416$ is the exact analog of our grade-$0$ seed; the numerical values differ due to vertex conventions (minimal-area vs.\ hyperbolic), but the algebraic structure is identical. The ghost dilaton decouples at cubic order by ghost number conservation, just as $V^{T,T,D} = 0$ in our level-$(0{+}2)$ vertex table.

At quartic order, the tachyon self-coupling $\ka^2\,V^{(4)}_0 = -3.0172\,t^4$~\cite{Belopolsky:1994sk,Moeller:2004yy} is large and negative. In the Yang--Zwiebach effective potential, this eliminates the cubic critical point when the tachyon is treated in isolation~\cite{Yang:2005rx}. In our framework, the corresponding statement is that the tachyon-only equation has only the trivial solution $g_T = 0$ (\S\ref{sec:grade-0}), and the seam-graded expansion around the cubic seed diverges badly ($\ep_B \simeq 2.6\times 10^{18}$). The B-term does not destroy the grade-$0$ seed per se --- the cubic fixed point $g_T^{(0)}$ remains a valid solution of the grade-$0$ equation --- but the grade-$1$ correction is $10^{18}$ times larger than the seed, and the formal series $\sum_n \Phi_n$ shows no sign of convergence.

The solution is restored by including the ghost dilaton. Yang and Zwiebach show that the $t^3 d$ quartic coupling forces the dilaton to condense ($d_* \simeq 0.49$), generating a compensating quintic term that restores the critical point at $t_* \simeq 0.338.$ In our framework, the dilaton enters through the multi-level grade-$0$ seed and the B-term at grade~$1.$ The detailed mechanism is left to the multi-level computation.

The seam-graded expansion and the Yang--Zwiebach level truncation are related by a dictionary:
\begin{center}
\renewcommand{\arraystretch}{1.3}
\begin{tabular}{l|l|l}
\hline
\textbf{Yang--Zwiebach} & \textbf{Seam-graded} & \textbf{Content} \\
\hline
Cubic ($V^{(3)}$) & Grade $0$ & Algebraic seed \\
Quartic ($V^{(4)}$) & Grades $\leq 1$ & B-term \\
$n$-point ($V^{(n)}$) & Grades $\leq n{-}3$ & Cumulative \\
\hline
\end{tabular}
\end{center}
The correspondence is not one-to-one: the A-term at grade~$0,$ when iterated, generates contributions to all interaction orders, and conversely the quartic vertex receives contributions from both the grade-$1$ B-term and the iterated A-term. The principal advantage of the seam-graded expansion is that it resums all interaction orders at each grade: one need not compute $V^{(4)},$ $V^{(5)},\ldots$ separately. A direct quantitative comparison between the two formulations requires matching vertex conventions, which we do not attempt here; the comparison that is meaningful is structural.

\section{Discussion}\label{sec:discussion}

\subsection{Convergence}\label{sec:convergence}

The convergence of the seam-graded expansion $\Phi = \sum_{n=0}^{\infty}\Phi_n$ is the central open question of this paper. We have three data points: a fully solvable toy model that converges, a scalar truncation that diverges, and the full multi-level system whose convergence is not yet determined.

In the stubbed scalar theory of \S\ref{sec:review-eq69}, every grade contributes a correction $\phi_n(t_1) = c_n\,\ue^{t_1\mu^2/2}$ whose coefficient $c_n$ is determined recursively. The grade-$0$ seed accounts for an exponentially suppressed fraction of the full solution; the remaining corrections must make up the deficit. As shown by F{\i}rat and Valdes-Meller~\cite{Firat:2024ajp}, the series converges to the exact nonperturbative solution $\phi(t_1) = (2\mu^4/\ka)\,\ue^{t_1\mu^2/2},$ and the mass spectrum $M^2 = \mu^2 > 0$ is obtained without resumming the potential. This confirms that the formal structure of the seam-graded expansion --- the grade-independent linearized operator, the recursive source evaluation, the resolvent $(I-M)^{-1}$ --- is compatible with convergence in at least one nontrivial example.

In the scalar tachyon sector of the full string theory, the situation is strikingly different. The grade-$1$ source satisfies $|S_1^T(\Lstar)|/|g_T^{(0)}(\Lstar)| \simeq 2.6\times 10^{18}$ (\S\ref{sec:S1-continuation}), meaning the first correction exceeds the seed by eighteen orders of magnitude. This enormous ratio is a direct consequence of the B-term integral, which probes the entire border-length range $\ell\in [0,\infty)$ and receives large contributions from the region where the vertex is exponentially enhanced. The grade-$1$ correction $g_T^{(1)}(\Lstar) = -S_1^T(\Lstar)$ therefore overwhelms the grade-$0$ seed, and there is no reason to expect subsequent corrections to be smaller. The tachyon-only expansion diverges.

This divergence is not unexpected: the tachyon-only equation has no nontrivial solution (\S\ref{sec:grade-0}), and a divergent series around a seed with no nearby fixed point is the natural consequence. The Yang--Zwiebach diagnosis (\S\ref{sec:comparison-YZ}) is that the ghost dilaton must condense alongside the tachyon.

In the multi-level system, the source $S_n(\Lstar)$ is a vector in Fock space whose components may exhibit inter-level cancellations absent in the tachyon-only truncation. Whether these cancellations are sufficient to achieve $\ep_B < 1$ is determined by the spectral radius of the linearized map at the correct multi-level seed: convergence requires this spectral radius to be strictly less than one in all directions transverse to the scaling mode $g^{(0)}.$ We note that there is no known structural reason to expect the required cancellations: the tachyon component of $S_1$ receives its dominant contribution from the essential singularity at small $\ell$ (governed by $h = -1$), and while massive states ($h > 0$) have convergent integrals individually, the inter-level cancellation would need to occur in the tachyon component after summing over all intermediate states in the B-term. Whether such cancellations arise is a purely empirical question that the multi-level grade-$1$ computation will answer.

Whether the multi-level seam-graded expansion converges, or is asymptotic and requires resummation, is an open question. If the series is asymptotic, the algebraic framework remains valid but would require resummation techniques to extract physical predictions.

\subsection{Special functions and resummation}\label{sec:special-functions}

The source integrals $S_n(L_1)$ are integrals of products of Gamma functions at imaginary arguments, integrated over border lengths. They belong to the broad class of Barnes-type integrals, though the non-factorized dependence on all three border lengths prevents direct application of standard identities. Whether $S_n$ evaluates to a known special function is an open mathematical problem; regardless, the integrals are explicit and computable to arbitrary precision.

The seam-graded expansion provides a formal series $\Phi(L) = \sum_{n=0}^{\infty}(I - M(L))^{-1}\,S_n(L).$ If the series is asymptotic rather than convergent, Borel resummation may apply. The large-$n$ behavior of $\|S_n\|,$ which is controlled by the geometry of moduli spaces $V_{0,n+3},$ determines whether the Borel transform has singularities on the positive real axis. Computing $S_n$ for the first several grades in the multi-level system will diagnose the convergence behavior.

\subsection{Quantum extension}\label{sec:quantum}

The seam-graded expansion is a classical ($\hbar = 0$) construction. The FVM topological recursion extends to higher genus~\cite{Firat:2024ajp}, with the nonseparating $D$-term providing loop corrections. The seam grading generalizes naturally: at genus~$1,$ the recursion produces the one-bordered torus vertex $V_{1,1}(L_1)$ from $V_{0,3}$ via the BV Laplacian. The physical payoff is the mass spectrum of fluctuations around the tachyon vacuum~\cite{Yang:2005rx}.

\subsection{Future directions}\label{sec:future}

Three directions follow from this work. First, the numerical computation should be pushed to higher level truncation and higher seam grade. The multi-level grade-$1$ source $S_1(\Lstar)$ will determine whether inter-level cancellations reduce $\ep_B$ below unity; if the expansion is asymptotic, it would serve as the starting point for resummation. Second, finding an algebraic approach to find the analytic soltuion is a central open problem. Third, the genus-$1$ extension gives access to the mass spectrum around the tachyon vacuum~\cite{Yang:2005rx}, the most direct physical observable.

\medskip

In summary, in the sector of zero-momentum Lorentz scalars --- closed under the equations of motion by Lorentz symmetry, though the physical vacuum may lie outside it --- the seam-graded expansion reduces the closed string tachyon vacuum problem to a sequence of matrix inversions with explicit source integrals. The algebraic structure is universal, the numerical verification through $15$ states shows no obstruction, and the grade-$1$ source is under control after analytic continuation. The decisive test --- whether inter-level cancellations tame the tachyon-only $\ep_B \simeq 10^{18}$ --- is the immediate next step. The deeper questions --- whether inter-level cancellations tame the divergence, and whether an analytic solution can be found --- are the most important open problems.

\section*{Acknowledgments}

We thank Atakan Hilmi F{\i}rat for discussions and detailed feedback on the manuscript. The author thanks Siddharth Mishra-Sharma for providing access to Claude Max, which was used extensively in the preparation of this work. The author thanks Venkatesa Chandrasekaran, Mehmet Demirtas, and Gowri Kurup for encouraging the author to explore AI for science. The work of MK is supported by Simons Investigator award (MPS-SIP-00507021).

\section*{Author contributions}

M.~Kim conceived the project, directed the research, and wrote the final manuscript. Claude (Anthropic, Claude Opus 4.6), accessed through the publicly available Claude Code interface, served as a computational and research assistant throughout the project, performing symbolic and numerical computations, adversarial review of intermediate results, literature analysis, and drafting of preliminary text under the author's close supervision. All physical ideas, key insights, interpretive judgments, and final decisions were made by M.~Kim.

\newpage
\appendix

\section{Explicit grade-by-grade formulas (tachyon sector)}\label{app:explicit}

For reference, we collect the tachyon projection at each grade after $\ka$ absorption, for general threshold length $L.$ In the tachyon-only sector, $\BG$ acts as the scalar $\mathrm{BG}$ (eq.~\eqref{eqn:BG on tachyon}), so the $\BG$-dressed component reduces to $[\BG(L)\,\Phi(L)]_T = \mathrm{BG}(L)\,g_T(L).$ We write the formulas in terms of the bare vertex $V_T$ and the scalar $\mathrm{BG}$ explicitly.

\textbf{Grade~$0$:}
\begin{align}
    g_T^{(0)}(L_1) &= C_A\,V_T(L_1,L,L)\;\big[\mathrm{BG}(L)\,g_T^{(0)}(L)\big]^2\,,\label{eqn:grade0 profile}\\[4pt]
    g_T^{(0)}(L) &= \frac{1}{C_A\,V_T^{\mathrm{eff}}(L,L,L)}\,,\label{eqn:grade0 value}\\[4pt]
    g_T^{(0)}(L_1) &= \frac{V_T^{\mathrm{eff}}(L_1,L,L)}{V_T^{\mathrm{eff}}(L,L,L)}\;g_T^{(0)}(L)\,,\label{eqn:grade0 ratio}
\end{align}
where $V_T^{\mathrm{eff}}(L_1,L_2,L_3) = V_T(L_1,L_2,L_3)\,\mathrm{BG}(L_2)\,\mathrm{BG}(L_3)$ as defined in \S\ref{sec:grade-0}.

\textbf{Grade~$1$ source:}
\begin{equation}
    S_1^T(L_1) = C_B\;\mathrm{BG}(L)\,g_T^{(0)}(L)\int_0^{\infty}\!\ud\ell\;\eR(L_1,L,\ell)\;V_T(L_1,L,\ell)\;\mathrm{BG}(\ell)\,g_T^{(0)}(\ell)\,.\label{eqn:grade1 source}
\end{equation}
Here $\mathrm{BG}(L)\,g_T^{(0)}(L) = [\BG(L)\,\Phi_0(L)]_T$ and $\mathrm{BG}(\ell)\,g_T^{(0)}(\ell) = [\BG(\ell)\,\Phi_0(\ell)]_T$ are the $\BG$-dressed tachyon components.

\textbf{Grade~$1$ correction:}
\begin{equation}
    g_T^{(1)}(L) = -S_1^T(L)\,,\qquad g_T^{(1)}(L_1) = S_1^T(L_1) + M_T(L_1)\;g_T^{(1)}(L)\,.\label{eqn:grade1 correction}
\end{equation}

\textbf{Grade~$2$ source (four terms):}
\begin{enumerate}[(2a)]
    \item C-term with $\Phi_0$ in both slots (grade $0+0+2=2$):
    \begin{equation}
        C_C\!\int_0^{\infty}\!\ud\ell_1\!\int_0^{\infty}\!\ud\ell_2\;\eD(L_1,\ell_1,\ell_2)\;V_T(L_1,\ell_1,\ell_2)\;\mathrm{BG}(\ell_1)\,g_T^{(0)}(\ell_1)\;\mathrm{BG}(\ell_2)\,g_T^{(0)}(\ell_2)\,.\label{eqn:grade2a}
    \end{equation}
    \item B-term with $\Phi_0$ at threshold length, $\Phi_1$ integrated (grade $0+1+1=2$):
    \begin{equation}
        C_B\;\mathrm{BG}(L)\,g_T^{(0)}(L)\int_0^{\infty}\!\ud\ell\;\eR(L_1,L,\ell)\;V_T(L_1,L,\ell)\;\mathrm{BG}(\ell)\,g_T^{(1)}(\ell)\,.\label{eqn:grade2b}
    \end{equation}
    \item B-term with $\Phi_1$ at threshold length, $\Phi_0$ integrated (grade $1+0+1=2$):
    \begin{equation}
        C_B\;\mathrm{BG}(L)\,g_T^{(1)}(L)\int_0^{\infty}\!\ud\ell\;\eR(L_1,L,\ell)\;V_T(L_1,L,\ell)\;\mathrm{BG}(\ell)\,g_T^{(0)}(\ell)\,.\label{eqn:grade2c}
    \end{equation}
    \item A-term quadratic in $\Phi_1$ (grade $1+1+0=2$):
    \begin{equation}
        C_A\,V_T(L_1,L,L)\;\big[\mathrm{BG}(L)\,g_T^{(1)}(L)\big]^2\,.\label{eqn:grade2d}
    \end{equation}
\end{enumerate}

\textbf{Grade~$2$ correction:} $g_T^{(2)}(L) = -S_2^T(L).$ The linearized operator at grade $2$ is $M$ (same as grade $1$), not a Fredholm integral operator.

\section{Key numerical values}\label{app:numerics}

All numerical values in this appendix are evaluated at the threshold length $L = \Lstar.$ All mapping radii are computed from the hypergeometric local coordinate~\eqref{eqn:local coordinate} by extracting $\rho_j = |f_j'(0)|$ to $50$-digit precision. The function $v$ in eq.~\eqref{eqn:v function} determines $\ue^{2\ui v}$ only modulo~$\pi;$ in the numerical implementation we fix the branch by requiring $\rho_j$ to agree with the hypergeometric result, which is equivalent to absorbing the integer $\tilde{l}_j$ from eq.~\eqref{eqn:mapping radius} into the branch choice. This convention sets $\tilde{l} = -1$ uniformly and shifts $v$ relative to the principal-branch evaluation of eq.~\eqref{eqn:v function} at asymmetric border lengths.

\begin{center}
\renewcommand{\arraystretch}{1.3}
\begin{tabular}{l|l|l}
\hline
Quantity & Value & Source \\
\hline
$\Lstar$ & $2\sinh^{-1}(1) = 1.76274717\ldots$ & \cite{Firat:2024ajp}  \\
$\lamstar$ & $\Lstar/(2\pi) = 0.28055\ldots$ & Definition \\
$\rhostar = \rho(\Lstar,\Lstar,\Lstar)$ & $0.03566\ldots$ & \cite{Firat:2021ukc}  \\
$V_T(\Lstar,\Lstar,\Lstar) = \rhostar^{-6}$ (bare) & $4.87\times 10^8$ & \cite{Firat:2021ukc} \\
$\mathrm{BG}(\Lstar) \equiv \mathrm{BG}(\Lstar,\Lstar,\Lstar)$ & $1.126$ & \S\ref{sec:review-bghost} \\
$V_T^{\mathrm{eff}}(\Lstar,\Lstar,\Lstar) = \mathrm{BG}(\Lstar)^2\,\rhostar^{-6}$ & $6.17\times 10^8$ & \S\ref{sec:grade-0} \\
$C_A$ & $2$ & \eqref{eqn:combinatorial coefficients} \\
$C_B$ & $4$ & \eqref{eqn:combinatorial coefficients} \\
$C_C$ & $2$ & \eqref{eqn:combinatorial coefficients} \\
$g_T^{(0)}(\Lstar)$ & $1/(C_A\,V_T^{\mathrm{eff}}) \simeq 8.10\times 10^{-10}$ & \S\ref{sec:grade-0} \\
$M_T(\Lstar)$ & $2$ (exact) & \S\ref{sec:grade-1} \\
$(I-M_T(\Lstar))^{-1}$ & $-1$ (exact) & \S\ref{sec:all-grade} \\
$S_1^T(\Lstar)$ & $(2.08\pm 0.05)\times 10^9$ & \S\ref{sec:S1-continuation} \\
$|S_1^T(\Lstar)|/|g_T^{(0)}(\Lstar)|$ & $\simeq 2.6\times 10^{18}$ & \S\ref{sec:S1-continuation} \\
\hline
\end{tabular}
\end{center}

\bibliographystyle{JHEP}
\bibliography{refs}

\end{document}